\documentclass[a4paper,11pt]{article}
\pdfoutput=1 

\usepackage{jheppub} 
\usepackage{epstopdf} 
\usepackage[T1]{fontenc} 
\usepackage[vcentermath]{youngtab}
\usepackage{bbm}

\usepackage{graphics}
\usepackage{inputenc}
\usepackage{xspace}
\usepackage{amsmath}
\usepackage{amssymb}
\usepackage{url}
\usepackage{rotating}
\usepackage{enumerate}
\usepackage{graphicx}
\usepackage{def} 

\title{\boldmath QCD multiplet bases with arbitrary parton ordering}

\author[]{Malin Sjodahl} 
\author[]{ and Johan Thor\'en}

\affiliation[]{Department of Astronomy and Theoretical Physics, Lund
  University, S{\"o}lvegatan 14A, 223\,62 Lund, Sweden}

\emailAdd{malin.sjodahl@thep.lu.se}
\emailAdd{johan.thoren@thep.lu.se}

\abstract{
  We develop an algorithm for recursively constructing orthogonal multiplet
  bases for the color space of QCD, for any order of partons and any $\Nc$.
  This recipe is then applied for explicitly constructing some
  of these bases.
  Using the bases, a corresponding set of Wigner $6j$ coefficients are
  calculated.
  The Wigner coefficients offer a method of using multiplet bases without
  resorting to the explicit expressions of the basis vectors,
  which lead to a significant speed-up compared to other methods of
  treating full color structure.
}

\begin{document} 
\preprint{LU-TP 18-26, MCNET-18-25}
\maketitle
\flushbottom

\section{Introduction}
\label{sec:introduction}
One challenge caused by the high multiplicity of colored particles at the LHC
is the complexity of the color space of QCD.
Traditionally this has been tackled with the use of non-orthogonal bases,
such as the trace bases 
\cite{Paton:1969je, Berends:1987cv, Mangano:1987xk,
  Mangano:1988kk, Kosower:1988kh, Nagy:2007ty, Sjodahl:2009wx, Alwall:2011uj,
  Sjodahl:2014opa, Platzer:2012np}
and color flow bases \cite{'tHooft:1973jz,Kanaki:2000ms,Maltoni:2002mq}.

These bases have several advantages, conceptual simplicity, simple
relations for gluon emission and gluon exchange,
etc. \cite{Mangano:1988kk,Nagy:2007ty,Sjodahl:2009wx}, but the non-orthogonality and
overcompleteness of the bases quickly become an issue when squaring amplitudes with many
colored partons. Multiplet bases \cite{Kyrieleis:2005dt,Dokshitzer:2005ig,Sjodahl:2008fz,Beneke:2009rj, Keppeler:2012ih, Du:2015apa, Sjodahl:2015qoa,Alcock-Zeilinger:2016cva}, based on $SU(\Nc)$
representations, are orthogonal and minimal, hence curing that issue,
but are not yet generally available.

Work on multiplet bases for an arbitrary number of partons was initiated in
\cite{Keppeler:2012ih} by giving a general recipe for constructing multiplet
bases, which in principle can be used for any number of quarks, antiquarks and
gluons. The bases, presented in \cite{Keppeler:2012ih}, have later been used
for calculating Wigner $6j$ coefficients \cite{Sjodahl:2015qoa}, which can be used to decompose
amplitudes into multiplet bases and for performing amplitude recursion in multiplet bases
\cite{Du:2015apa}.

\begin{figure}[tbp]
  \label{fig:basis_comparison}
  \centering
  \includegraphics[width=0.28\textwidth]{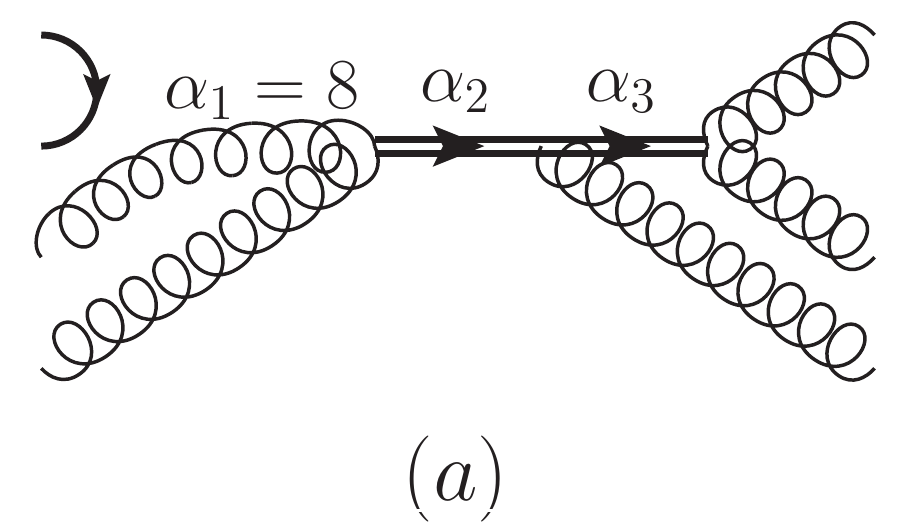}
  \includegraphics[width=0.28\textwidth]{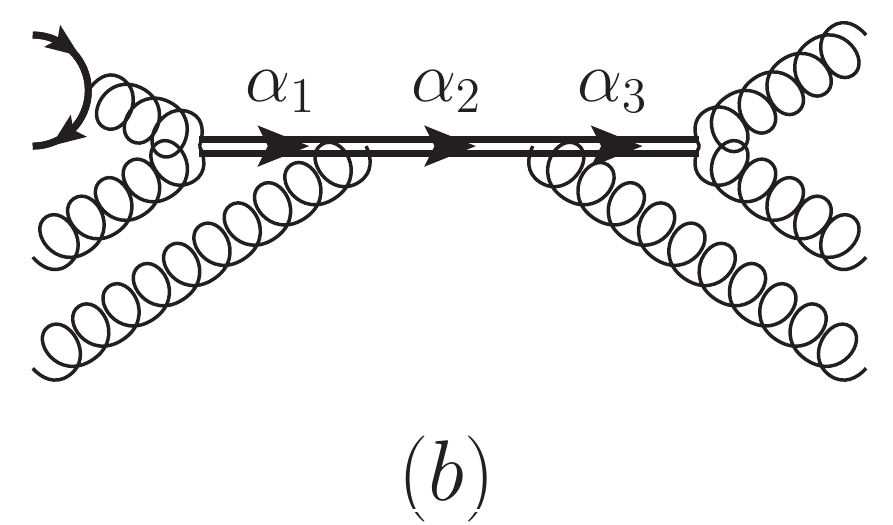}
  \includegraphics[width=0.336\textwidth]{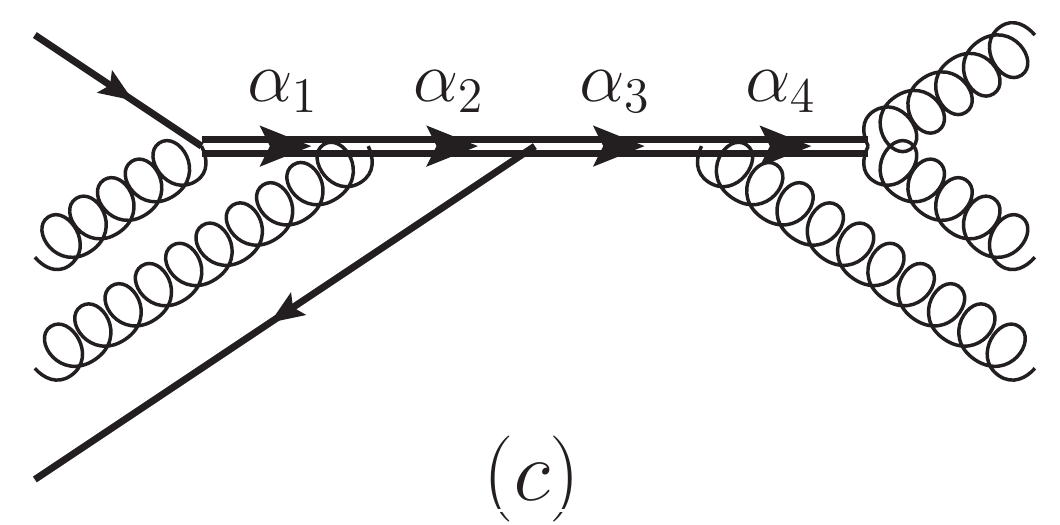}
  \caption{Basis vectors with quark-antiquark pairs combined into a singlet, (a),
    an octet (b) and a basis vector where they are not combined, (c).
    Double-lines carry the representation given by their label.
    The basis vectors are divided into an incoming side, with half of the partons (rounded up)
    and an outgoing side, with the other half (rounded down), swapping a quark (antiquark)
    from the incoming to the outgoing side (or vice versa) only changes its label to
    an antiquark (quark).
  }
\end{figure}

In this article we generalize the basis construction from \cite{Keppeler:2012ih},
by allowing arbitrary groupings of quark, antiquark and gluon representations.
The recipe in \cite{Keppeler:2012ih} constructs basis vectors
of the form (a) and (b) in \figref{fig:basis_comparison}, where the quarks and antiquarks
are grouped into singlets and octets\footnote{
  We occasionally refer to representations using their $\Nc=3$
  dimension, although clearly the dimension differs for other values of $\Nc$.
}. The new construction introduced in this paper
avoids this requirement, and allows 
basis vectors of the form
in (c), where the quarks and antiquarks attach
directly to a chain of ``backbone'' representations.
Using the new basis vectors, new Wigner $6j$ coefficients that can occur
in the basis decomposition are calculated,
in a similar
manner as in \cite{Sjodahl:2015qoa}.
With the algorithm in this paper,
more
general multiplet bases can thus be constructed and used for color decomposition.
This allows for choosing a more
appropriate multiplet basis for the problem at hand. This is particularly useful
for applications of the multiplet basis in a recursion or parton shower context,
as the choice of multiplet basis affects the efficiency significantly \cite{Du:2015apa}.
We remark, however, that all projectors and basis vectors considered here still have all
external partons, quarks, antiquarks and gluons, attached to a ``backbone'' chain of
general representations, $\alpha_1$,\dots,$\alpha_n$ in \figref{fig:basis_comparison}.
Lifting this condition, yet more general basis constructions can be imagined.

This paper is organized as follows: In \secref{sec:using_wigner6j} 
we recapitulate the method of \cite{Sjodahl:2015qoa,cvi08} for decomposing scalar products into
multiplet bases using
Wigner $6j$ coefficients,
and in \secref{sec:orthogonality} the construction history method of \cite{Keppeler:2012ih}
for achieving transversality for the $SU(\Nc)$ projectors is outlined.
In \secref{sec:notation}, an $\Nc$-independent notation for $SU(\Nc)$ representations
is introduced.
Section \ref{sec:Projectorsqqbar}
introduces the recursive method of constructing projectors with one
additional quark (or antiquark) from known projectors with any parton content.
The construction from \cite{Keppeler:2012ih} for projectors with an additional gluon is
briefly summarized in \secref{sec:ProjectorsGluon}.
Then we show how to use the projectors to construct multiplet basis vectors in
\secref{sec:basis} and in turn how to evaluate Wigner $6j$ coefficients with the vectors
in \secref{sec:6j}. Finally, we conclude in \secref{sec:conclusions}.

\section{Using Wigner $6j$ coefficients for color structure decomposition}
\label{sec:using_wigner6j}
In this paper we make use of the birdtrack notation \cite{cvi08} for the group theoretical
calculations. We will frequently use the completeness relation
\begin{eqnarray}
  \label{eq:CRDiagrammatic}
  \raisebox{-0.325\height}{
    \includegraphics[scale=0.45]{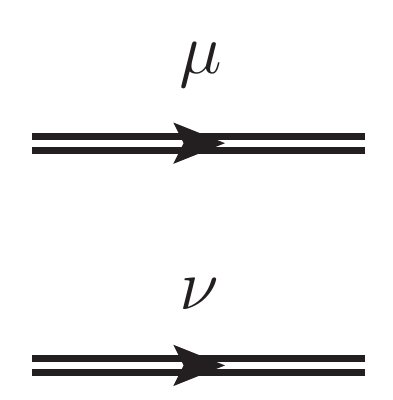}
  }
  =
  \sum_{\alpha\in\mu\otimes\nu}{
    \frac{d_\alpha}{\includegraphics[scale=0.3]{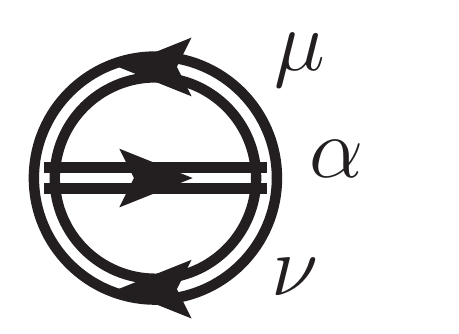}\hspace{-1.5mm}}
    \raisebox{-0.325\height}{
      \includegraphics[scale=0.45]{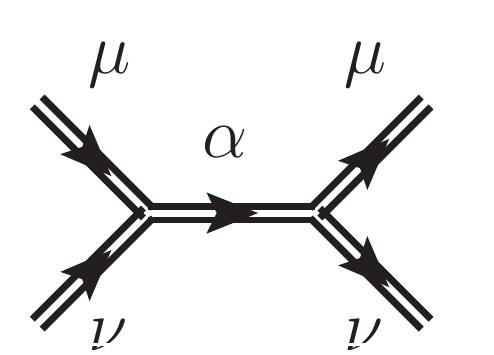}
    }
  },
\end{eqnarray}
where Greek letters denote arbitrary representations and
the denominator is a group invariant, called a Wigner $3j$ coefficient.
We will also use Schur's lemma
\begin{equation}\label{eq:SchursLemma}
\raisebox{-0.39\height}{
\includegraphics[scale=0.45]{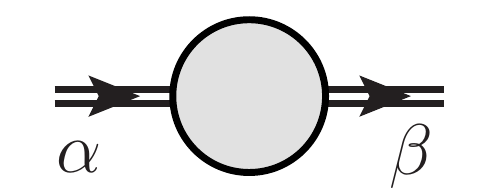}
}
\hspace{-2mm}
=
\frac{
\hspace{-2mm}
\raisebox{-0.4\height}{
\includegraphics[scale=0.45]{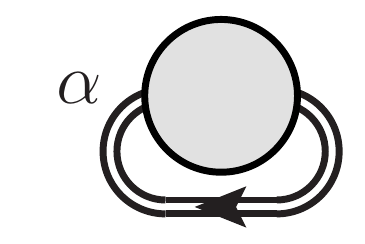}
}
\hspace{-2mm}
}{
d_\alpha
}
\delta^{\alpha}_{\;\beta}
\raisebox{-0.45\height}{
\includegraphics[scale=0.45]{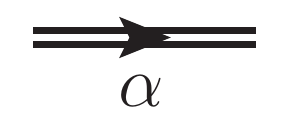}
},
\end{equation}
and the vertex correction relation
\begin{equation}\label{eq:VertexCorrection}
\raisebox{-0.44\height}{
	\includegraphics[scale=0.4]{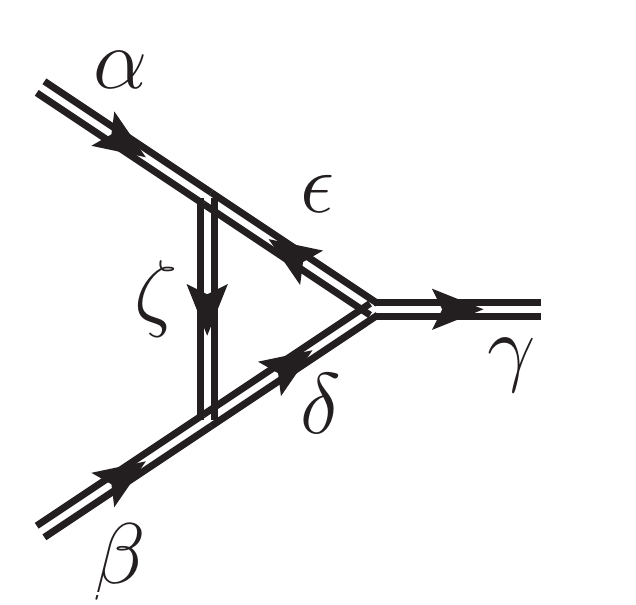}
}
\hspace{-2mm}
=
\sum_{a}
{
\frac{
	\raisebox{-0.45\height}{
		\includegraphics[scale=0.4]{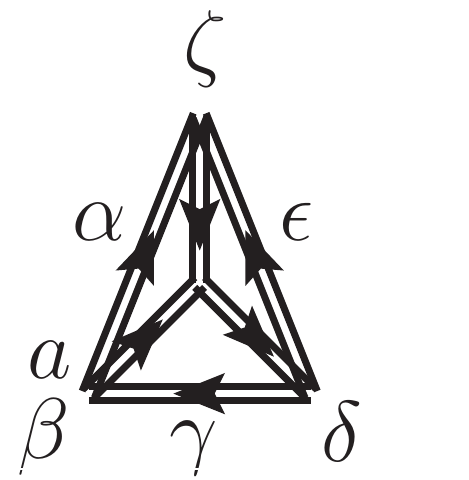}
	}
	\hspace{-3mm}
}{
	\raisebox{-0.45\height}{
		\includegraphics[scale=0.3]{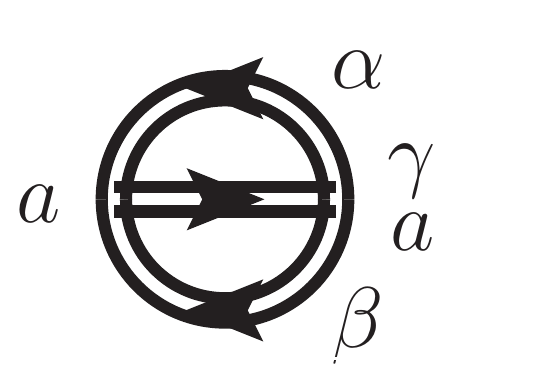}
	}
	\hspace{-3mm}
}
\raisebox{-0.43\height}{
  \includegraphics[scale=0.4]{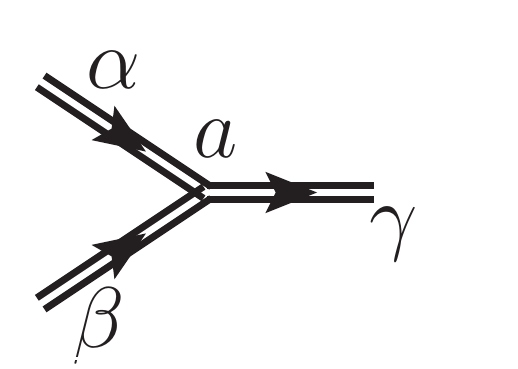}
}
}
\hspace{-3mm},
\end{equation}
where the numerators in the ratios 
are group invariants,
so-called Wigner $6j$ coefficients, 
and the sum over $a$ runs over every instance\footnote{
  If a tensor product contains more than one instance of a representation
  an additional label $a$ has to be used to distinguish them,
  for example in
  $8\otimes{}8=1\oplus{}8\oplus{}8\oplus{}10\oplus{}\overline{10}\oplus{}27$
  ($A\otimes{}A$ for $\Nc=3$) the two copies of the octet have to be
  distinguished. These vertex labels are implicit in many places in this paper
  where they are not used.
} of $\gamma$ in
$\alpha\otimes\beta$. For readability the representation labels of the inner
representations in the $6j$ coefficients ($\beta$, $\delta$ and $\zeta$
in \eqref{eq:VertexCorrection})
are placed by the corners.

\begin{figure}[tbp]
  \label{fig:ScalarProduct}
  \centering
  \includegraphics[scale=0.4]{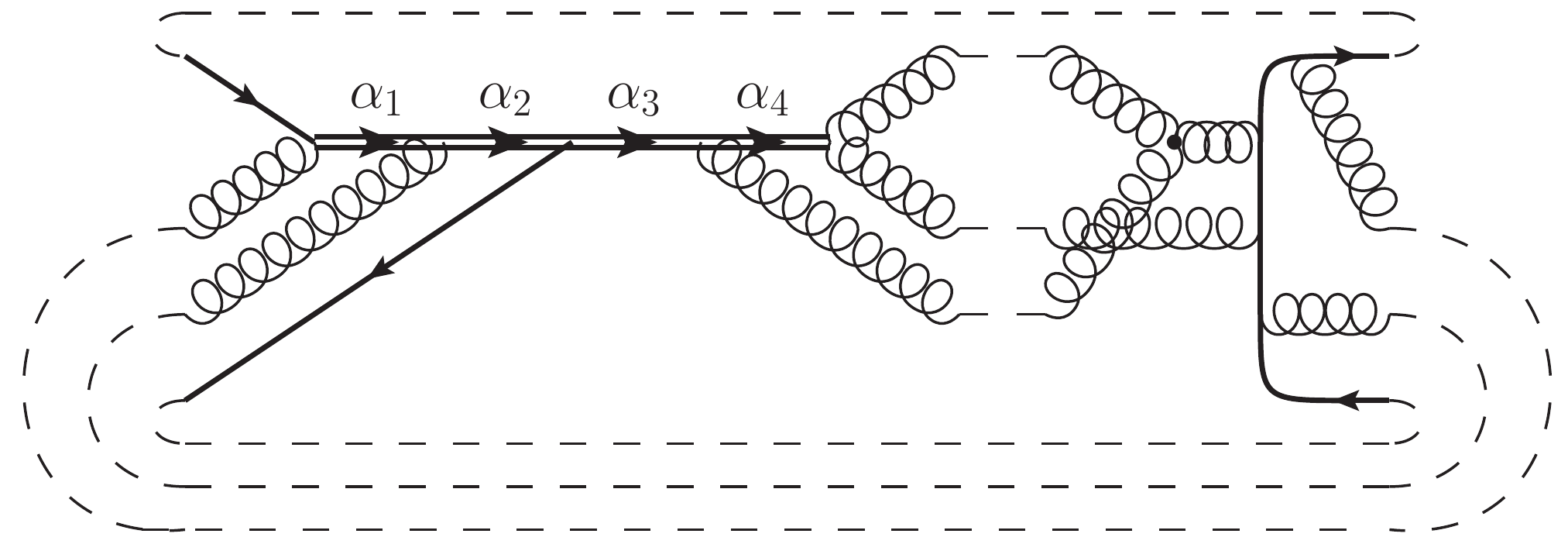}
  \caption{Example of a scalar product between a basis vector and a
    color structure, the dashed lines indicate how the external color indices
    should be contracted.}
\end{figure}

For QCD we are interested in color summed/averaged quantities, due to confinement.
Hence, we are concerned with evaluating scalar products of color structures.
If $\mathbf{c}_1$ and $\mathbf{c}_2$ are two color structures, the scalar product
is defined to be
\begin{equation}
  \label{eq:ScalarProductIndex}
  \langle{}c_1|c_2\rangle = \sum_{a_1,a_2,\dots}{c_1^{*a_1a_2\dots}c_2^{a_1a_2\dots}},
\end{equation}
where $a_i=1,\dots\Nc$ if parton $i$ is a quark and $a_i=1,\dots,\Nc^2-1$ if $i$ is
a gluon.
In the birdtrack notation, the scalar product is a fully contracted vacuum bubble,
where the conjugated color structure has been mirrored and all representation arrows
have changed direction, an example is shown in \figref{fig:ScalarProduct}.

As described in \cite{cvi08} in general, and in \cite{Sjodahl:2015qoa} for QCD in particular,
fully contracted color structures can be decomposed by repeatedly using
completeness relations,
Schur's lemma
and
the vertex correction relation (eqs.~(\ref{eq:CRDiagrammatic}-\ref{eq:VertexCorrection}))
on loops, resulting in a vacuum bubble with fewer vertices.
Any fully contracted color structure must, clearly, contain loops.
For the scalar product between a basis
vector, of the form considered here (i.e.~with a backbone of general representations),
and a leading order amplitude, e.g.~as in \figref{fig:ScalarProduct}, loops
where all of the vertices in the loop are from the backbone
of general representations, except for one vertex
coming from the color structure to decompose
can always be found.
For such loops,
there will only be two necessary types of steps in the contraction of the
loop\footnote{
  For next-to-leading order, another type of loop would need to be considered,
  which is studied in detail in \cite{Sjodahl:2015qoa}. However, this does
  not require Wigner $6j$ coefficients of another form, but with other
  constraints on the representations in the coefficient.
} \cite{Sjodahl:2015qoa}.
The first step in reducing such a loop, is
\begin{equation}\label{eq:CRApplicationLoopType1}
  \raisebox{-0.4\height}{
    \includegraphics[scale=0.45]{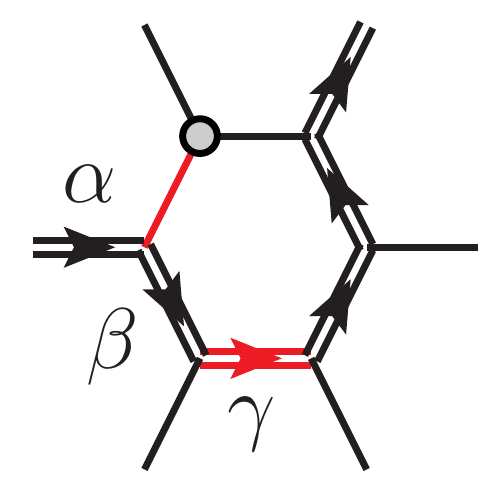}
  }
  =
  \sum_{\psi}{
    \frac{d_{\psi}}{
      \hspace{0.5mm}
      \raisebox{-0.45\height}{
	\includegraphics[scale=0.4]{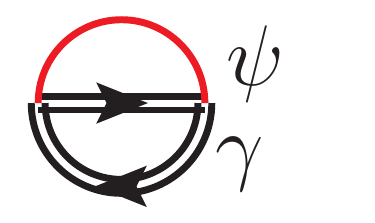}
      }
      \hspace{-3mm}
    }
    \raisebox{-0.4\height}{
      \includegraphics[scale=0.45]{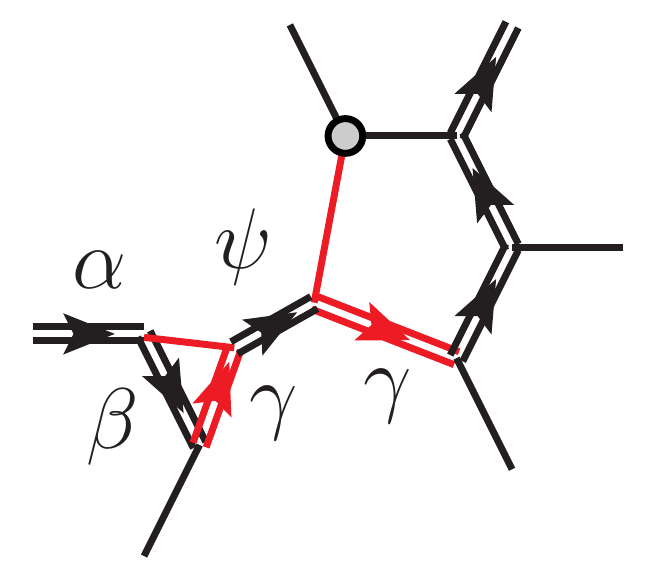}
    }
  }
=
\sum_{\psi,a}{
		\frac{d_{\psi}}{
		\hspace{0.5mm}
		\raisebox{-0.45\height}{
			\includegraphics[scale=0.4]{Figures/Representations/Wig3jCR}
		}
		\hspace{-3mm}
	}
	\frac{
		\hspace{-1mm}
		\raisebox{-0.1\height}{
			\includegraphics[scale=0.4]{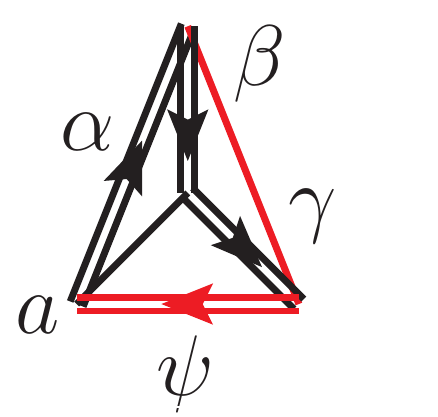}
		}
		\hspace{-5mm}
	}{
		\hspace{0.5mm}
		\raisebox{-0.45\height}{
			\includegraphics[scale=0.4]{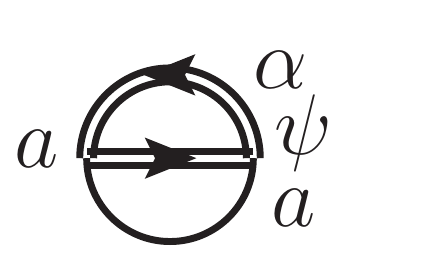}
		}
		\hspace{-4mm}
	}
\raisebox{-0.4\height}{
	\includegraphics[scale=0.45]{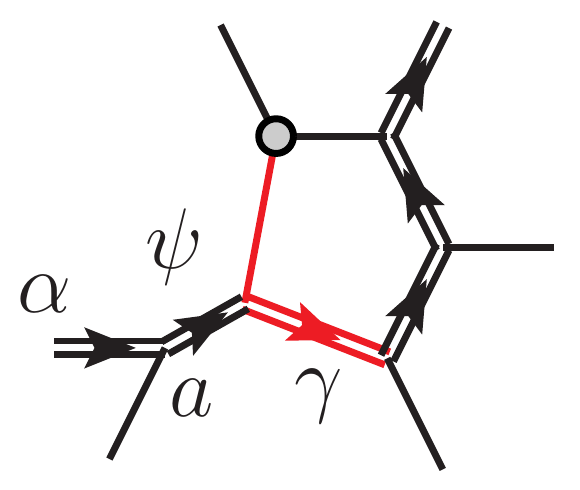}
}
}
,
\end{equation}
where single lines without arrows can carry the fundamental or the adjoint
representation and
$\raisebox{-0.25\height}{\includegraphics[scale=0.4]{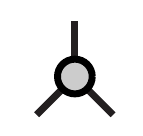}}$
is the vertex from the color structure to decompose,
either the antisymmetric triple gluon vertex,
$\raisebox{-0.25\height}{\includegraphics[scale=0.4]{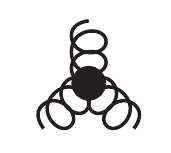}}$,
or
the quark-gluon vertex,
$\raisebox{-0.25\height}{\includegraphics[scale=0.4]{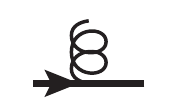}}$.
(We remark that there will often be smarter ways of contracting indices
where completeness relations are avoided, but the
above contraction can \emph{always} be performed.)
This step can be applied to a loop of any length, and noting that the right hand side
is of similar form, but
contains a shorter loop than the left hand side, it is clear that this procedure will achieve large parts of the
basis decomposition, when applied repeatedly.
In the final step, only three vertices remain in the loop, and a vertex correction can be used to get
\begin{equation}\label{eq:CRApplicationLoopType1LastCR}
  \raisebox{-0.49\height}{
    \includegraphics[scale=0.4]{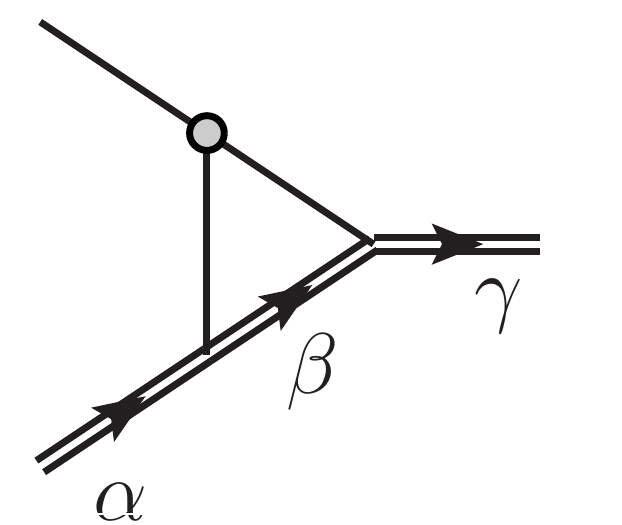}
  }
  \hspace{-3mm}
  =
  \sum_{a}{
    \frac{
      \hspace{-0.5mm}
      \raisebox{-0.1\height}{
	\includegraphics[scale=0.4]{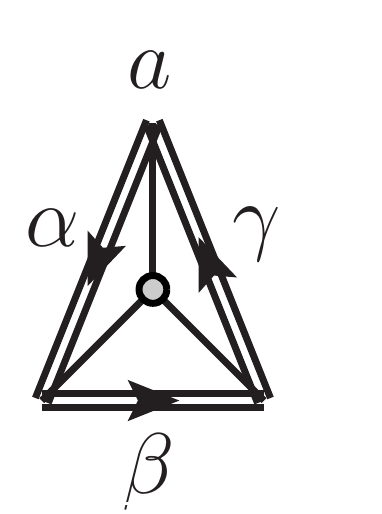}
      }
      \hspace{-4mm}
    }{
      \hspace{-0.5mm}
      \raisebox{-0.45\height}{
	\includegraphics[scale=0.4]{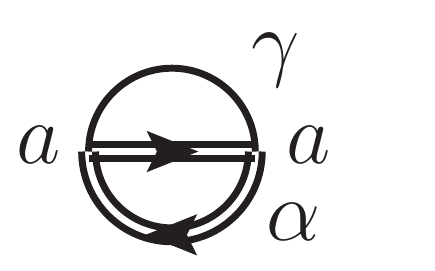}
      }
      \hspace{-4mm}
    }
    \raisebox{-0.5\height}{
      \includegraphics[scale=0.4]{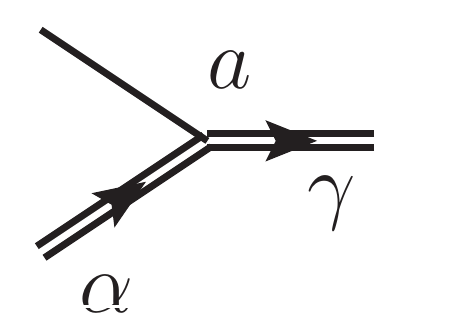}
    }
  }
  \hspace{-3mm}
  .
\end{equation}
We note that \eqref{eq:CRApplicationLoopType1} and \eqref{eq:CRApplicationLoopType1LastCR}
only require Wigner $6j$ coefficients of two specific forms.
By considering all possible ways of assigning the thin lines to be in the triplet,
antitriplet or octet representation, the Wigner $6j$ coefficients required to
decompose color structures into the basis vectors considered in this paper,
where the quarks, antiquarks
and gluons can be in any order, are therefore of the form:
\begin{equation}\label{eq:Wigner6j4Representations}
\raisebox{-0.45\height}{
	\includegraphics[scale=0.4]{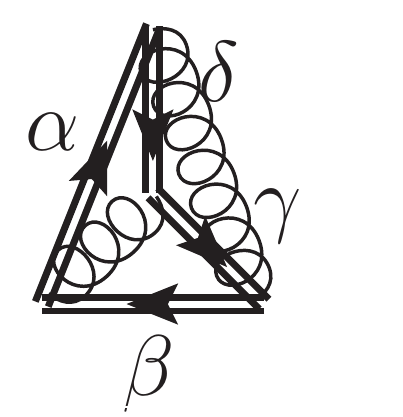}
}
\hspace{-4mm},\;
\raisebox{-0.45\height}{
	\includegraphics[scale=0.4]{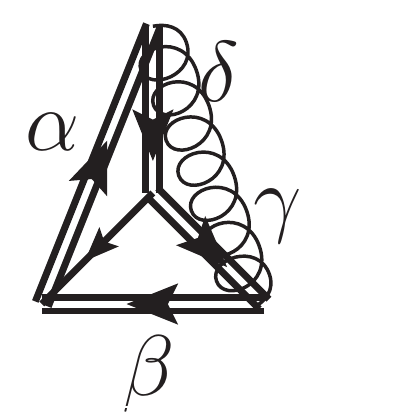}
}
\hspace{-4mm},\;
\raisebox{-0.45\height}{
	\includegraphics[scale=0.4]{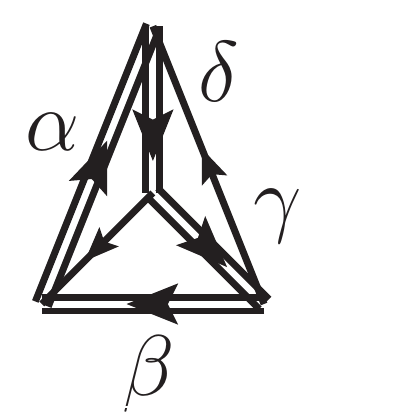}
}
\hspace{-3mm}
\end{equation}
and
\begin{equation}\label{eq:Wigner6j3Representations}
\raisebox{-0.45\height}{
	\includegraphics[scale=0.4]{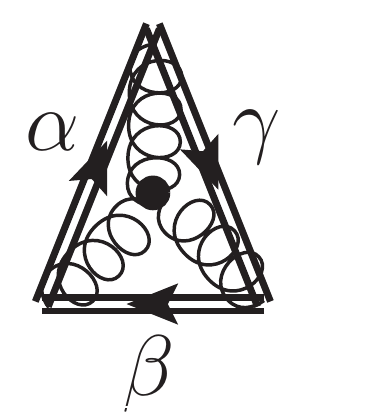}
}
\hspace{-4mm},\;
\raisebox{-0.45\height}{
	\includegraphics[scale=0.4]{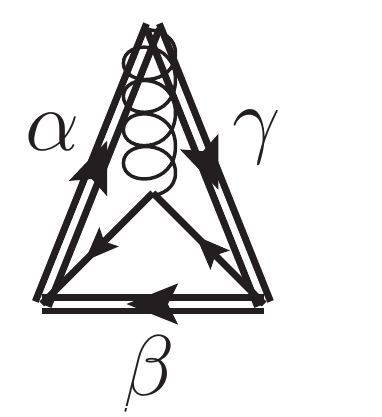}
}
\hspace{-4mm},
\end{equation}
where symmetries of the coefficients have been used to write down the minimal set
of required coefficients.
Compared to the coefficients that were needed in \cite{Sjodahl:2015qoa},
the second and third
coefficients in \eqref{eq:Wigner6j4Representations}
and the second coefficient in \eqref{eq:Wigner6j3Representations} are new\footnote{
  In \cite{Sjodahl:2015qoa} an additional type Wigner $6j$ coefficient
  containing the symmetric triple gluon vertex
  was required, due to how the quarks were treated, but it is no longer
  required here since QCD does not contain symmetric triple gluon
  vertices.
  The required coefficient was of the form of the first coefficient in
  \eqref{eq:Wigner6j3Representations}, with the antisymmetric triple gluon vertex
  in the middle exchanged for a symmetric triple gluon vertex.
}.
There will also be new allowed representations
for $\alpha$, $\beta$, $\gamma$ and $\delta$, in \eqref{eq:Wigner6j4Representations}
and \eqref{eq:Wigner6j3Representations},
as compared to \cite{Sjodahl:2015qoa}, since the representations are
no longer required to be in $\Adj^{\otimes{}n_g}$, where $\Adj$ is the
adjoint representation.

The more general multiplet bases introduced in this paper, allow for more Wigner
$6j$ coefficients to be calculated. Using these for the basis decomposition
for processes involving quarks avoids the complicated decomposition
in \cite{Sjodahl:2015qoa} (appendix A.2).
Constructing more general multiplet bases is especially useful for applications such
as recursion relations with fermions or for full color parton showers \cite{Platzer:2012np,Nagy:2012bt,Sjodahl:2014opa,Nagy:2015hwa,Platzer:2018pmd,Isaacson:2018zdi}. Requiring a certain
grouping of the quarks and antiquarks will often result in expressions with more terms.
In the recursion in \cite{Du:2015apa}, the pure gluon basis vectors had a gluon
order chosen to minimize
the number of terms required. In a generalization of that work to fermion recursion,
a similar ordering of the partons, minimizing the number of terms,
would be possible with the basis vectors introduced here.
As the color structures encountered for a parton shower are similar to those encountered in
recursion, due to the iterative way of adding one emission at the time, the additional freedom
in the choice of multiplet basis ordering is equally useful in a parton shower context.

\section{Transversality through construction history}
\label{sec:orthogonality}
Transversality of two projectors, $P_\alpha$ and $P_\beta$, is defined as
$P_\alpha P_\beta=\delta_{\alpha\beta}P_\alpha$, where we have used
idempotency, $P_\alpha P_\alpha = P_\alpha$.
As described in \cite{Sjodahl:2008fz, Keppeler:2012ih} the projection operators may be constructed
to be transversal by successively combining partons into representations in a so-called
construction history. One way of achieving this is by combining the partons as shown in
the following equation
\begin{equation}\label{eq:ConstructionHistory}
  \raisebox{-0.4\height}{
    \includegraphics[scale=0.45]{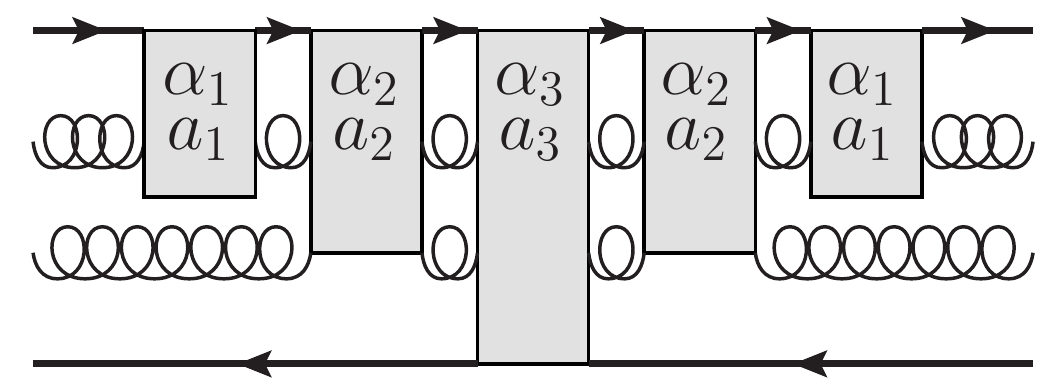}
  } \Leftrightarrow
  \raisebox{-0.4\height}{
    \includegraphics[scale=0.45]{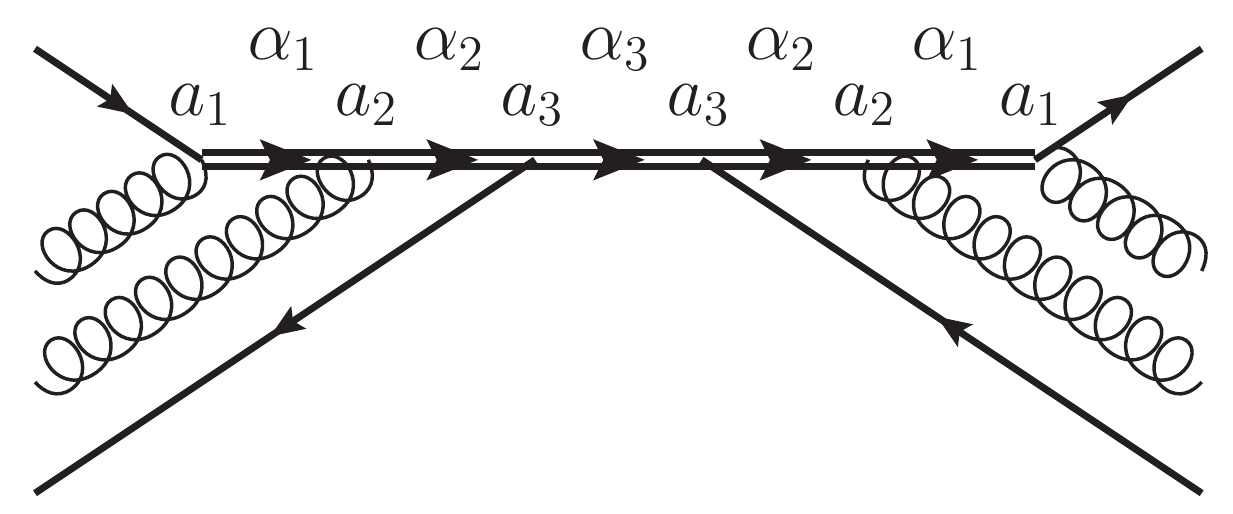}
  }.
\end{equation}
Here the rectangles with two labels, e.g.\ $\alpha_1$ and $a_1$ for the leftmost square, denote the instance
$a_1$ of the projector $\alpha_1$.
Hence, in the construction history used in this paper, the representations are combined
two at a time, from top to bottom. In \eqref{eq:ConstructionHistory}, the first quark
and gluon are thus combined into a representation $\alpha_1\in{}\V\otimes{}\Adj$, where
$\V$ is the fundamental representation ($\Vc$ is the complex conjugate of the fundamental
representation).
This is then combined with the next parton (in this case another gluon) into a
representation $\alpha_2\in{}\alpha_1\otimes{}\Adj$, etc.\ until all partons on the left
side are combined into one specific representation, here $\alpha_3$. For transversal
projectors, the contraction of two projectors, of the form in \eqref{eq:ConstructionHistory},
would correspond to
\begin{equation}\label{eq:ProjectorContraction}
  P_\alpha{}P_\beta{}
  =
  \raisebox{-0.5\height}{
    \includegraphics[scale=0.45]{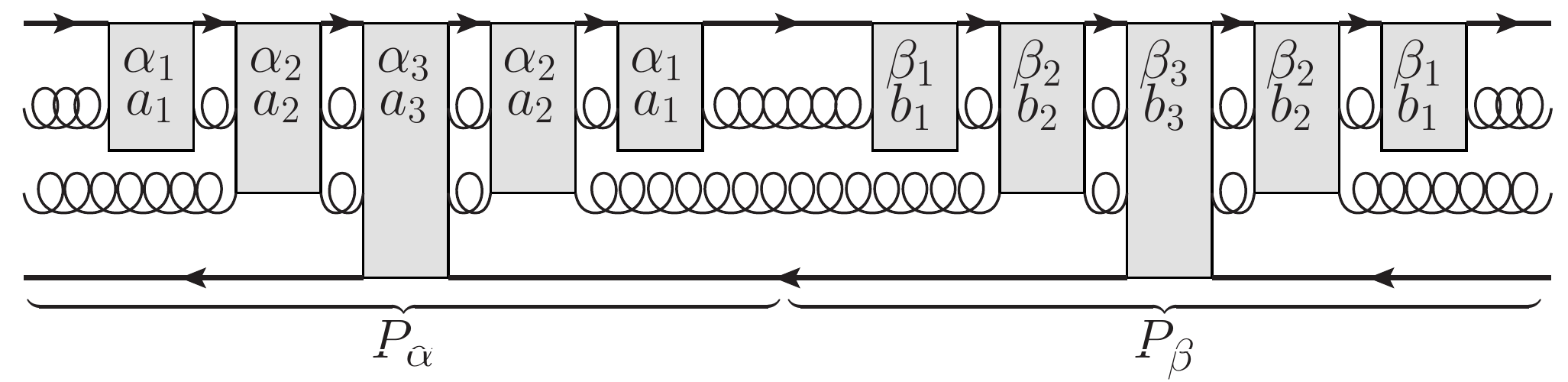}
  },
\end{equation}
which vanishes, by Schur's lemma, \eqref{eq:SchursLemma}, for all possible combinations
of representations and vertices, unless the two projectors are identical,
i.e.~$\alpha_i=\beta_i$ and $a_i=b_i$ for $\forall{}i$. If the representations are the same,
but correspond to different instances (e.g.~$\alpha_i=\beta_i$ but $a_i\neq{}b_i$ for some $i$)
one can always construct the projectors such that they are transversal. For our purposes,
starting in a representation $M$,
this will only happen in $M\otimes{}\Adj$, and not in $M\otimes{}\V$ or $M\otimes{}\Vc$,
since in $M\otimes{}\V$ or $M\otimes{}\Vc$ each representation can only appear once.

The transversality of the projectors will, as in \cite{Keppeler:2012ih}, be used
to construct vectors that are orthogonal
under the scalar product in the color space, \eqref{eq:ScalarProductIndex} and
\figref{fig:ScalarProduct} in birdtrack notation.

\section{$\Nc$-independent Young tableau notation}
\label{sec:notation}
A representation $M$ for the group $SU(N_c)$ is associated with a Young
diagram which, in general, depends on the number of colors, $\Nc$, see
e.g. \cite[secs. 7.12 \& 10]{Ham62}.
To achieve a basis construction in an $\Nc$-independent way, we need an $\Nc$-independent
labeling of representations, which will be introduced in this section.
This notation is used to perform the tensor products of
arbitrary $SU(\Nc)$ representations with $\V$, $\Vc$ and $\Adj$.
Every representation is in our notation associated with a quark diagram and a barred
antiquark diagram \footnote{In a late stage of this work, it was pointed out to us that
  the same idea is used already in \cite{doi:10.1063/1.1665059}.},
\begin{equation}
\label{eq:QuarkAndAntiquarkDiagramsGeneral}
\left(Q,\overline{\tilde{Q}}\right).
\end{equation}
As an example consider
\begin{equation}
\label{eq:QuarkAndAntiquarkDiagrams}
\left(\yng(2),\overline{\yng(3,1)}\right),
\end{equation}
where the left Young diagram is the quark diagram and the right, barred,
diagram is the antiquark diagram. For a specific $\Nc$, the Young diagram this
corresponds to, is obtained by conjugating the barred diagram and then
merging it with the quark diagram. For the rest of this paper we
will refer to the right diagram of \eqref{eq:QuarkAndAntiquarkDiagramsGeneral} and
\eqref{eq:QuarkAndAntiquarkDiagrams} as the barred diagram and the, equivalent, conjugated
diagram (which is an $\Nc$ specific diagram) as the conjugated antiquark diagram.
The conjugation denoted by the bar is performed by removing the bar, marking the boxes,
rotating them $180^\circ$, then adding boxes on top of each column until
each column is $\Nc$ boxes high and finally remove the marked boxes, see \cite[sec. 9.8]{cvi08}.
For the example, if $\Nc=3$, the conjugated antiquark diagram is ${\tiny\yng(3,2)}$, as can
be seen from
\begin{equation}
  \label{eq:ConjugatingDiagram}
  \young(\bullet\bullet\bullet,\bullet)
  \rightarrow
  \young(\hfil\hfil\hfil,\hfil\hfil\bullet,\bullet\bullet\bullet)
  \rightarrow
  \begin{matrix}
    \begin{sideways}
      \Nc-1
    \end{sideways}
    \hspace{1.5mm}
    \begin{sideways}
      \Nc-1
    \end{sideways}
    \hspace{1.5mm}
    \begin{sideways}
      \Nc-2
    \end{sideways}
    \\
    \yng(3,2)
    \\
    \phantom{1}
  \end{matrix}
  \hspace{1mm}.
\end{equation}
The merging of the diagrams in \eqref{eq:QuarkAndAntiquarkDiagrams} is
done by aligning the first rows and
putting the diagrams next to each other with the conjugated antiquark diagram
to the left. For the example in \eqref{eq:QuarkAndAntiquarkDiagrams},
we attach the quark diagram 
to the right of the
conjugated antiquark diagram, \eqref{eq:ConjugatingDiagram}, i.e.
\begin{equation}
  \label{eq:QuarkAndAntiquarkDiagramsExample}
  \begin{matrix}
    \hspace{-0.5mm}
    \begin{sideways}
      \Nc-1
    \end{sideways}
    \hspace{1.5mm}
    \begin{sideways}
      \Nc-1
    \end{sideways}
    \hspace{1.5mm}
    \begin{sideways}
      \Nc-2
    \end{sideways}
    \hspace{1.5mm}
    1
    \hspace{2.1mm}
    1\\
    \yng(5,2)
  \end{matrix}
  \hspace{1mm},
\end{equation}
where the numbers above each column indicate the length of the column for a general
$\Nc$. For low enough $\Nc$, the resulting Young diagram might not be an
admissible diagram, due to the rightmost column in the conjugated antiquark
diagram being shorter than the leftmost column of the quark diagram. 
For the example, the chosen $\Nc$ is the lowest $\Nc$ for which the diagram
is admissible, since for $\Nc=2$ the third column of the
conjugated antiquark diagram has zero boxes.
In general, a representation will be present for $\Nc$ larger
than or equal to the length of the leftmost column of the barred diagram
plus the length of the leftmost column of the quark diagram.

This notation is introduced to handle tensor products between $\V$, $\overline{V}$
and $\Adj$ and a general SU($\Nc$) representation in a $\Nc$-independent
way. The tensor product with the fundamental representation can be divided into
a part where a box is added to the quark diagram and a part where a box is
removed from the barred diagram.
To see this, we introduce a notation for tensor products with quarks
\begin{equation}
\label{eq:QuarkAndAntiquarkDiagramsTensorProductNotationGeneral}
\left(Q,\overline{\tilde{Q}}\right)\otimes{}\V
=
\left(Q\otimes{}\V,\overline{\tilde{Q}}\right)
\oplus
\left(Q,\overline{\tilde{Q}}``\otimes{}"\V \right)
,
\end{equation}
in our case
\begin{equation}
\label{eq:QuarkAndAntiquarkDiagramsTensorProductNotation}
\left(\yng(2),\overline{\yng(3,1)}\right)\otimes{}\V
=
\left(\yng(2)\otimes{}\V,\overline{\yng(3,1)}\right)
\oplus
\left(\yng(2),\overline{\yng(3,1)}``\otimes{}"\V\right)
.
\end{equation}
The second 
term has quotation marks to indicate that it is not exactly a tensor product, the
exact meaning of it will be explained and motivated below.
Starting from \eqref{eq:QuarkAndAntiquarkDiagramsExample}, the reason for why
this notation can be used to uniquely describe Young tableau multiplication for
general $\Nc$ will be shown. The tensor product of the representation
\eqref{eq:QuarkAndAntiquarkDiagramsExample} with $\V$ is 
\begin{equation}
  \label{eq:NormalTensorProduct}
  \begin{matrix}
    \hspace{-0.5mm}
    \begin{sideways}
      \Nc-1
    \end{sideways}
    \hspace{1.5mm}
    \begin{sideways}
      \Nc-1
    \end{sideways}
    \hspace{1.5mm}
    \begin{sideways}
      \Nc-2
    \end{sideways}
    \hspace{1.5mm}
    1
    \hspace{2.1mm}
    1\\
    \yng(5,2)
  \end{matrix}
  \hspace{1mm}
  \begin{matrix}
    \\
    \otimes
  \end{matrix}
  \hspace{1mm}
  \begin{matrix}
    1\\
    \yng(1)
  \end{matrix}
  \begin{matrix}
    \\
    =
  \end{matrix}
  \begin{matrix}
    \hspace{-0.5mm}
    \begin{sideways}
      \Nc-1
    \end{sideways}
    \hspace{1.5mm}
    \begin{sideways}
      \Nc-1
    \end{sideways}
    \hspace{1.5mm}
    \begin{sideways}
      \Nc-2
    \end{sideways}
    \hspace{1.5mm}
    1
    \hspace{2.1mm}
    1
    \hspace{2.1mm}
    1\\
    \yng(6,2)
  \end{matrix}
  \hspace{1mm}
           \begin{matrix}
             \\
             \oplus
           \end{matrix}
           {\color{lightgray}
           \hspace{1mm}
           \begin{matrix}
             \hspace{-0.5mm}
             \begin{sideways}
               \Nc-1
             \end{sideways}
             \hspace{1.5mm}
             \begin{sideways}
               \Nc-1
             \end{sideways}
             \hspace{1.5mm}
             \begin{sideways}
               \Nc-2
             \end{sideways}
             \hspace{1.5mm}
             2
             \hspace{2.1mm}
             1\\
             \yng(3,2)\hspace{-.1mm}\yng(2,1)
           \end{matrix}
         }
         \hspace{1mm}
         \begin{matrix}
           \\
           \oplus
         \end{matrix}
         \hspace{1mm}
  \begin{matrix}
    \hspace{-0.5mm}
    \begin{sideways}
      \Nc
    \end{sideways}
    \hspace{1.5mm}
    \begin{sideways}
      \Nc-1
    \end{sideways}
    \hspace{1.5mm}
    \begin{sideways}
      \Nc-2
    \end{sideways}
    \hspace{1.5mm}
    1
    \hspace{2.1mm}
    1\\
    \yng(5,2,1)
  \end{matrix}
  \hspace{1mm}
  \begin{matrix}
    \\
    \oplus
  \end{matrix}
  \hspace{1mm}
  \begin{matrix}
    \hspace{-0.5mm}
    \begin{sideways}
      \Nc-1
    \end{sideways}
    \hspace{1.5mm}
    \begin{sideways}
      \Nc-1
    \end{sideways}
    \hspace{1.5mm}
    \begin{sideways}
      \Nc-1
    \end{sideways}
    \hspace{1.5mm}
    1
    \hspace{2.1mm}
    1\\
    \yng(5,3)
  \end{matrix}
  \hspace{1mm}.
\end{equation}
For $\Nc=3$ the second, grayed out, diagram is clearly not allowed, since the rows are not
left-justified. However, if $\Nc\geq4$, then the third column of the diagram is $\geq2$
boxes high and it is then equal to or higher than the fourth column.
The first two diagrams of \eqref{eq:NormalTensorProduct} correspond one-to-one to the two
allowed ways of adding a box to the quark diagram in
\eqref{eq:QuarkAndAntiquarkDiagrams}. The corresponding statement is true for
general representations, $(Q,\overline{\tilde{Q}})$ as well, since all admissible ways
of adding a box to the
quark diagram correspond directly to a diagram in the tensor product of the
representation with $\V$, once the possibly disallowed diagram (where the box
is placed in the leftmost column of the quark diagram) has been crossed
out if $\Nc$ is too small.
If $\Nc$ is high enough, the diagram coming from adding the box to the first
column in the quark diagram will always be allowed.

That the two last diagrams in \eqref{eq:NormalTensorProduct} come from the two admissible ways of
removing a box from the barred diagram in
\eqref{eq:QuarkAndAntiquarkDiagrams} is possibly less clear. Considering the tensor
product of the conjugated antiquark diagram with $\V$, but leaving the diagram in the form
of the middle diagram of \eqref{eq:ConjugatingDiagram} such that it is easy to
see which antiquark diagram gives each of the possible representations, we find
\begin{equation}
\label{eq:TensorProductWithAntiquarkDiagram}
\begin{matrix}
	\hspace{-0.5mm}
	\begin{sideways}
		\Nc-1
	\end{sideways}
	\hspace{1.5mm}
	\begin{sideways}
		\Nc-1
	\end{sideways}
	\hspace{1.5mm}
	\begin{sideways}
		\Nc-2
	\end{sideways}
\\
	\young(\hfil\hfil\hfil,\hfil\hfil\hfil,\hfil\hfil\bullet,\bullet\bullet\bullet)
\end{matrix}
\otimes\yng(1)
=
\begin{matrix}
	\hspace{-0.5mm}
	\begin{sideways}
		\Nc
	\end{sideways}
	\hspace{1.5mm}
	\begin{sideways}
		\Nc-1
	\end{sideways}
	\hspace{1.5mm}
	\begin{sideways}
		\Nc-2
	\end{sideways}
\\
	\young(\hfil\hfil\hfil,\hfil\hfil\hfil,\hfil\hfil\bullet,\hfil\bullet\bullet)
\end{matrix}
\hspace{0.5mm}\oplus
\begin{matrix}
	\hspace{-0.5mm}
	\begin{sideways}
		\Nc-1
	\end{sideways}
	\hspace{1.5mm}
	\begin{sideways}
		\Nc-1
	\end{sideways}
	\hspace{1.5mm}
	\begin{sideways}
		\Nc-1
	\end{sideways}
\\
	\young(\hfil\hfil\hfil,\hfil\hfil\hfil,\hfil\hfil\hfil,\bullet\bullet\bullet)
\end{matrix}
\hspace{0.5mm}
\oplus
\begin{matrix}
	\hspace{-0.5mm}
	\begin{sideways}
		\Nc-1
	\end{sideways}
	\hspace{1.5mm}
	\begin{sideways}
		\Nc-1
	\end{sideways}
	\hspace{1.5mm}
	\begin{sideways}
		\Nc-2
	\end{sideways}
	\hspace{1.5mm}
	1
\\
	\young(\hfil\hfil\hfil\hfil,\hfil\hfil\hfil,\hfil\hfil\bullet,\bullet\bullet\bullet)
\end{matrix}
\hspace{1mm}.
\end{equation}
The first two representations correspond exactly to removing one box from the
barred diagram (in an admissible way). The third diagram would correspond to
adding $\Nc-1$ boxes to the barred diagram, and is the
difference between a proper tensor product and the ``$\otimes$''-product
in \eqref{eq:QuarkAndAntiquarkDiagramsTensorProductNotation}. It is omitted from
the tensor product in quotation marks as it would add a column of length one in
the middle of the combined Young diagram (as by the definition of the notation the
quark diagram should be merged to the right of the conjugated antiquark
diagram). The only case where it would lead to an admissible Young diagram after
the merge is if the quark diagram is a singlet (i.e. has no boxes), but for this
case that admissible Young diagram is already accounted for in the tensor product
with the quark diagram and should not be counted twice.
Hence we should never include this diagram, since it is either an inadmissible
diagram or it is already accounted for in the tensor product with the quark
diagram.
In total we have for our example
\begin{align}
\label{eq:QuarkAndAntiquarkDiagramsTensorProduct}
\left(\yng(2)\hspace{1mm},\overline{\yng(3,1)}\right)\otimes{}\V
&=
\left(\yng(2)\otimes{}\V,\overline{\yng(3,1)}\right)
\oplus
\left(\yng(2)\hspace{1mm},\overline{\yng(3,1)}``\otimes{}"\V\right)
\nonumber
\\
&=
\left(\yng(3)\hspace{1mm},\overline{\yng(3,1)}\right)
\oplus
\left(\yng(2,1)\hspace{1mm},\overline{\yng(3,1)}\right)
\nonumber
\\
&\oplus
\left(\yng(2)\hspace{1mm},\overline{\yng(2,1)}\right)
\oplus
\left(\yng(2)\hspace{1mm},\overline{\yng(3)}\right),
\end{align}
where line two corresponds to the first bracket on the right hand side of the first line, and the third line corresponds to the second bracket. 

A general tensor product will be of the
same form, terms corresponding to removing a box from the barred diagram and terms
corresponding to adding a box to the quark diagram.
From this we conclude that the multiplication introduced in 
\eqref{eq:QuarkAndAntiquarkDiagramsTensorProductNotationGeneral}
reproduces the result of $M\otimes \V$ for any $M$.
The case of $M \otimes \Vc$ can clearly be dealt with analogously.
Multiplication with $\Vc$ can then similarly as for $\V$ be seen
as representations coming from adding a box to the barred diagram or
removing a box from the quark diagram.

We have thus seen that the notation introduced in
\eqref{eq:QuarkAndAntiquarkDiagramsGeneral} along with the
multiplication in
\eqref{eq:QuarkAndAntiquarkDiagramsTensorProductNotationGeneral}
reproduces the result of Young tableau decomposition.
It can therefore be used to label SU($\Nc$) representations for general $\Nc$. 

For gluon multiplication we can use that
\begin{equation}\label{eq:qtimesqbar}
  \V \otimes{} \Vc = 1 \oplus{} \Adj,
\end{equation}
i.e. gluon multiplication can be treated as multiplication with a quark and an antiquark,
if we remove the contribution from the singlet.
From what we have shown above, the resulting representations of $M\otimes\V$, for any
representation $M$, can be seen as all ways of adding a box to the quark diagram
and all ways of removing a box from the barred diagram.
Analogously for $M\otimes\Vc$, the resulting representations correspond to
all ways of removing a box from the quark diagram and adding a box to the barred diagram.
Hence, we can divide $M\otimes{}\V\otimes{}\Vc$,
where $M = (Q,\overline{\tilde{Q}})$ is a general $SU(\Nc)$ representation, into different cases.
Either both of the boxes from $\V$ and $\Vc$ are altering the quark diagram $Q$
(or the antiquark diagram $\overline{\tilde{Q}}$) or one is altering $Q$ and the
other $\overline{\tilde{Q}}$.
For the first case we can further divide into two categories, such that we have
in total:
\begin{enumerate}[(i)]
\item The box from $\V$ is removed by the barred box from $\Vc$,
  meaning that both the quark and the antiquark act on either $Q$ or $\overline{\tilde{Q}}$,
  \begin{equation}
    \label{eq:GluonCaseSameRep}
    (Q\otimes{}\V``\otimes{}"\Vc,\overline{\tilde{Q}})
    \text{ or }
    (Q,\overline{\tilde{Q}}``\otimes{}"\V\otimes{}\Vc).
  \end{equation}
  This is always possible
  in at least one way, the quark box is added to the first row of the quark diagram
  and then removed by the antiquark box.
  In general the boxes can cancel in up to $\Nc$ different ways \cite{Keppeler:2012ih},
  out of which one way corresponds to the
  additional singlet in $\V\otimes{}\Vc$ and should be removed when considering
  the tensor product with $\Adj$.
  
\item The $\V$ box and the $\Vc$ barred box both act on $Q$ (or $\overline{\tilde{Q}}$),
  but do not cancel each other out,
  \begin{equation}
    \label{eq:GluonCaseSameFirstOccurrence}
    (Q\otimes{}\V``\otimes{}"\Vc,\overline{\tilde{Q}})
    \text{ or }
    (Q,\overline{\tilde{Q}}``\otimes{}"\V\otimes{}\Vc).
  \end{equation}
  This results in a different representation than
  $M$, but with the same number of quark boxes and barred boxes. Graphically this
  corresponds to moving one box within $Q$, or a barred box within $\overline{\tilde{Q}}$.
\item The $\V$ box acts on $\overline{\tilde{Q}}$ and the $\Vc$ barred box on $Q$,
  \begin{equation}
    \label{eq:GluonCaseLowerFirstOccurrence}
    (Q``\otimes{}"\Vc,\overline{\tilde{Q}}``\otimes{}"\V).
  \end{equation}
  This results in a representation different from $M$ with one less quark box
  and one less barred box.
\item The $\V$ box acts on $Q$ and the $\Vc$ barred box on $\overline{\tilde{Q}}$,
  \begin{equation}
    \label{eq:GluonCaseHigherFirstOccurrence}
    (Q\otimes{}\V,\overline{\tilde{Q}}\otimes{}\Vc).
  \end{equation}
  This results in a representation different from $M$ with one more quark box
  and one more barred box.
\end{enumerate}

\subsection{Equivalence of representations for finite $\Nc$}
\label{sec:finite_Nc}
In constructing the basis vectors for each overall representation, every
construction history of the left side has to be combined with every construction history of the
right side. In general there will be representations which are different for a high enough
$\Nc$, but are equivalent for a specific $\Nc$.
This will only occur for representations with different quark and antiquark
diagrams if some of the antiquark boxes are exchanged for
quark boxes (or vice versa). One example of this is
\begin{equation}
  \label{eq:finiteNcExample}
  \begin{matrix}
    \\
    \left(\yng(3),\overline{\yng(2)}\right)
  \end{matrix}
  \begin{matrix}
    \\
    \overset{\Nc=\hspace{0.25mm}3}{=}
  \end{matrix}
  \hspace{1mm}
  \begin{matrix}
    \hspace{-0.5mm}
    \begin{sideways}
      \Nc-1
    \end{sideways}
    \hspace{1.5mm}
    \begin{sideways}
      \Nc-1
    \end{sideways}
    \hspace{1.5mm}
    1
    \hspace{2.1mm}
    1
    \hspace{2.1mm}
    1\\
    \yng(5,2)
  \end{matrix}
  \hspace{1mm},
\end{equation}
for $\Nc=3$, which gives the same representation as the example,
\eqref{eq:QuarkAndAntiquarkDiagrams}, i.e. the representation in 
\eqref{eq:QuarkAndAntiquarkDiagramsExample}. This occurs because for a given
$\Nc$, the totally antisymmetric tensor with $\Nc$ fundamental indices is an
invariant, meaning that for finite $\Nc$, quark boxes can be traded for anti-quark boxes. In the example in \eqref{eq:finiteNcExample} two antiquark boxes are exchanged for one quark box.

This complication will not be present for perturbative QCD, precisely because the
$\epsilon$-tensor never appears in QCD vertices. Constructing multiplet bases that
do not preserve baryon number would require the addition of $\epsilon$-tensors in
the construction of the basis vectors, and would make the bases $\Nc$ specific.

\section{Constructing projectors with quarks and antiquarks}
\label{sec:Projectorsqqbar}
In this section we address the construction of the transversal projectors
required to decompose the space of $N_p$ partons and one quark (or antiquark).
The construction is recursive and assumes that a set of projectors for $N_p$ partons
and less are already known. The case of constructing projectors for $N_p$ partons
and one gluon can be dealt with as in \cite{Keppeler:2012ih}, where projectors for
$\Adj^{\otimes{}n_g}$ were constructed. Together
with the method presented in this section, projectors for any order of partons can be
constructed. As in \cite{Keppeler:2012ih} transversality of projectors is achieved
through the construction 
histories of the projectors.

The construction will, analogously to the recipe in \cite{Keppeler:2012ih}, use the concept of
``first occurrence'' to classify projectors into projectors with representations that
have already been encountered, and new representations.
The projectors corresponding to previously encountered representations
can be constructed by combining projectors that are, by assumption, already known.

In order to classify the representations occurring in $SU(\Nc)$ we thus need to
define the concept of first occurrence. 
In \cite{Keppeler:2012ih}, the first occurrence of a representation $M$,
is defined as the smallest integer $n_f$ such that
$M\in{}A^{\otimes{}n_f}$. For the construction of projectors with quarks, antiquarks and gluons this 
concept is here generalized into two integers, \nfq{} and \nfqbar{}, such that the first occurrence of
a multiplet is defined by the lowest numbers \nfq{} and \nfqbar{}, such that 
\begin{equation}
\label{eq:FirstOccurrence_q_qbar}
M\in{}\V^{\otimes{}\nfq}\otimes{}\Vc^{\otimes{}\nfqbar}.
\end{equation}
Since $\V\otimes{}\Vc = 1\oplus{}\Adj$ a representations in $\Adj^{\otimes{}n_g}$ for some 
number of gluons, $n_g$, with some first occurrence $n_f$, will have equal quark
and antiquark first occurrences, i.e.\ $\nfq=\nfqbar\equiv{}n_f$. In the quark and
antiquark diagram notation, as in \eqref{eq:QuarkAndAntiquarkDiagrams}, the
quark first occurrence is the number of boxes in the quark diagram and the
antiquark first occurrence is the number of boxes in the barred diagram. It is
then easy to see that the example in \eqref{eq:QuarkAndAntiquarkDiagrams} has
$\nfq=2$ and $\nfqbar=4$.

It is convenient to define 
\begin{equation}
\label{eq:FirstOccurrenceVector}
\vec{n}_f(M)=(\nfq(M),\nfqbar(M)),
\end{equation}
when considering the first occurrences of representations resulting from tensor
products with quarks or antiquarks. As mentioned in \secref{sec:finite_Nc}, for a
specific $\Nc$, some of the resulting representations can be equivalent
to representations with fewer quarks and antiquarks,
but throughout this paper we take
$\vec{n}_f(M)$ to have this $\Nc \to \infty$
meaning.

For a representation $M$ with
$\vec{n}_f(M)=(i,j)$, the representations in $M\otimes{}\V$ will have either
$\vec{n}_f=(i+1,j)$ (if the box is added to the quark diagram) or $\vec{n}_f=(i,j-1)$
(if a box is removed from the barred diagram), where the second type is only
present if $j>0$. This follows directly from the fact that $\nfq$ ($\nfqbar$)
is given by the number of boxes in the quark diagram (barred diagram). For multiplication
with an antiquark we analogously get either $\vec{n}_f(M')=(i,j+1)$ or
$\vec{n}_f(M')=(i-1,j)$, where $M'\in{}M\otimes{}\bar{q}$.

To construct the projectors with $\vec{n}_f=(i,j)$
the recursion requires projectors with lower first occurrence, defined as:
$\vec{n}_f(M')=(m,n)$
is lower than
$\vec{n}_f(M)=(i,j)$
if
$m\leq{}i$,
$n\leq{}j$, and at least one of $m$ and $n$ is lower than $i$ and $j$, respectively.
Higher first occurrence is then naturally defined as the opposite,
if $\vec{n}_f(M')$ is lower than $\vec{n}_f(M)$,
then $\vec{n}_f(M)$ is higher than $\vec{n}_f(M')$.
Representations which do not
fall into either of these categories (i.e. $n<i$ and $m>j$ or $n<i$ and $m>j$)
do not occur in the construction and will not be considered further. In the
construction, the representations will be classified based on their first occurrence.
The representations in $M\otimes{}\V$ that have
$\vec{n}_f=(\nfq(M)+1,\nfqbar(M))$ are, by the above definition, representations
with higher first occurrence, and the representations with
$\vec{n}_f=(\nfq(M),\nfqbar(M)-1)$ have lower first occurrence. Note that no
representation with the same first occurrence appears in the tensor product of
$M\otimes{}\V$ ($M\otimes{}\Vc$), only representations with lower or higher first
occurrence occur, whereas $M\otimes\Adj$ contains representations with higher,
lower and equal first occurrences.

For the $M\otimes\V$ and $M\otimes\Vc$ case, the projectors in the vector space of
the $N_p$ partons plus one quark will be
given by the representations in the tensor product of each projector for $N_p$
partons with the fundamental representation, $\V$. This accounts for every possible instance
of each representation, and the construction history will ensure transversality 
(that the new projectors are of the correct form will be shown in the
coming subsections). The case of $M\otimes\Adj$ is briefly discussed in
\secref{sec:ProjectorsGluon}.

The construction of the projectors will be divided into three cases, based on the difference 
in first occurrence with respect to the previous representation in the construction history 
and the shape of the Young diagram corresponding to the representation. 
The first type to be constructed is when the representation has a lower first occurrence, 
this is treated in \secref{subsec:ProjectorsLowerFirstOccurrence}. For the second type, 
where the first occurrence has increased, the construction is divided into two
cases. First, in \secref{subsec:ProjectorsFromTwoTensorProducts}, we consider a
standard case in which uniqueness of the new projector can be inferred by
building up the representation in two different ways. Then, in
\secref{subsec:ProjectorsFromOneTensorProduct}, we address a special case where
the quark (antiquark) diagram has rectangular shape, implying that the total
representation can only come from the multiplication of one single
representation with $\V$ ($\Vc$).

\subsection{Projectors with lower first occurrence}
\label{subsec:ProjectorsLowerFirstOccurrence}
In the tensor product of a representation $M$ with a quark (antiquark) there can be 
representations $M'$ where the antiquark (quark) first occurrence of $M'$ is lower than 
for $M$, corresponding to removing a barred box from the barred diagram
(box from the quark diagram).
The only case when there is no such representation $M'$ is when $\nfqbar(M)=0$ 
($\nfq(M)=0$), since in this case there are no barred boxes (boxes) to remove.
The corresponding projectors can be constructed by sandwiching the projector for
$M'$ (which is known by assumption) for fewer partons between the tensor product
of $M$ and a quark (antiquark), as follows
\begin{equation}
\label{eq:ProjectorLowerFirstOccurrence}
P_{M'}
\propto
\raisebox{-0.4\height}{
    \includegraphics[scale=0.45]{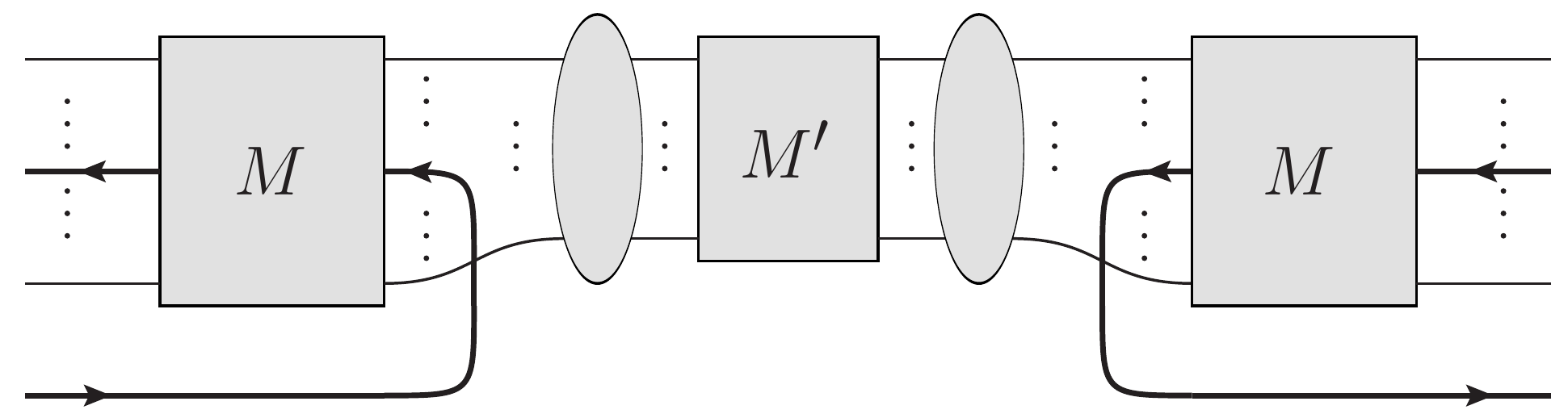}
    },
\end{equation}
where the gray blobs represent any connection of lines that makes the
expression non-vanishing. The projector onto the representation $M$ can have any
combination of partons, but there must be at least one antiquark (quark) if the added parton is a quark
(antiquark), or at least one gluon that can be split into a quark-antiquark
pair (if not, there cannot be any representation $M'$ with lower
$\nfqbar$ ($\nfq$)).

With respect to the equivalence of representations for finite $\Nc$,
we note that adding a box to the conjugated antiquark diagram is
always possible and will always result in a representation with lower $\nfqbar$.
Finite $\Nc$ thus never requires any special treatment in the
construction of projectors with lower first occurrence.

\subsection{Projectors with higher first occurrence}
\subsubsection{Construction from different tensor products}
\label{subsec:ProjectorsFromTwoTensorProducts}
In this section we will first prove that the tensor products  
$M_{1}\otimes{}\V$ and $M_{2}\otimes{}\V$ for two general representations
$M_1$ and $M_2$ can share at most one representation. We will then
show how to use this to construct projectors where $\nfq$ is increased
by one, i.e. the projectors corresponding to the first part of
\eqref{eq:QuarkAndAntiquarkDiagramsTensorProductNotationGeneral}. 
The case of addition of an antiquark $M \otimes{}\Vc$ where $\nfqbar$
is increased by one can clearly be treated analogously.

For this section we will consider an example representation,
$M'\in{}M\otimes\V$, for which
we wish to construct the projector. The argument is valid for any
$M'$ where the quark diagram is not rectangular, the remaining representations,
with rectangular quark diagrams,
will be dealt with in \secref{subsec:ProjectorsFromOneTensorProduct}.
For now we take $M'$ to be 
\begin{equation}
\label{eq:TwoTensorProducts}
M'=
\left(
\young(\hfil\hfil\hfil\hfil\star,\hfil\hfil\hfil\star,\hfil\star,\hfil,\star),
\overline{\tilde{Q}}
\right),
\end{equation}
where we have placed stars on every outer corner 
in the diagram
on the bottom-right side. In general we require $M'$ to have at least two
stars, implying that the quark diagram cannot be rectangular,
but we have no requirements on the barred diagram, $\overline{\tilde{Q}}$.
The stars mark the places where we can remove a box, giving a diagram, $M$, with quark
first occurrence reduced by one unit. Note that $M'\in{}M\otimes{}\V$,
since we will get the representation $M'$ by adding the box from $\V$ to $M$ exactly at
the spot where we removed a box to find $M$.
Now consider two different representations, found by removing a starred box from $M'$,
$M_1$ and $M_2$, for the example in \eqref{eq:TwoTensorProducts}, there are $(4\cdot3)/2$
different such choices. One possible choice is
\begin{equation}
\label{eq:TwoTensorProductsExample}
M_1=
\left(
\young(\hfil\hfil\hfil\hfil,\hfil\hfil\hfil\hfil,\hfil\hfil,\hfil,\hfil),
\overline{\tilde{Q}}
\right),
\;\;
M_2=
\left(
\young(\hfil\hfil\hfil\hfil\hfil,\hfil\hfil\hfil,\hfil\hfil,\hfil,\hfil),
\overline{\tilde{Q}}
\right),
\end{equation}
where $M_1$ is the representation where the starred box in the top row has been removed, 
and $M_2$ is the representation where the box in the second row has been removed.

We will prove that only the representation, $M'$, appears in both
of the tensor products $M_1\otimes{}\V$ and $M_2\otimes{}\V$. To see this,
we first consider the
representations corresponding to removing a barred box from the barred diagram
($\overline{\tilde{Q}}$) in $M_1\otimes\V$ or $M_2\otimes\Vc$.
These sets of representations from
$M_1\otimes{}\V$ and $M_2\otimes{}\V$
can clearly not have any overlap
, since their quark diagrams are different.
The representations corresponding to adding a box to the quark diagram will,
by construction have $M'$ in both tensor products. However, no other representation can
be in both tensor products, since they are missing one box each, at different places.
Thus adding a box cannot make them equivalent, except if it is added such that the
resulting representation is $M'$. This is true in general, independent of the shape
of the quark diagram. Below we will use that $M'$ is the only representation contained
in both $M_1\otimes\V$ and $M_2\otimes\Vc$ to construct the corresponding projector.

To construct the projector, we thus sandwich one of the tensor products in between the other tensor product, 
\begin{align}
\label{eq:TwoTensorProductsBirdtrack}
\raisebox{-0.4\height}{
    \includegraphics[scale=0.45]{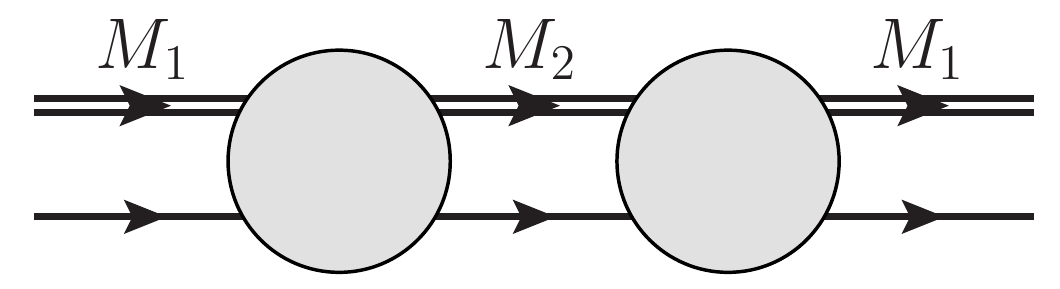}
}
& =
\sum_{\psi_1,\psi_2,\psi_3}{
  \frac{d_{\psi_1}}{
  \includegraphics[scale=0.3]{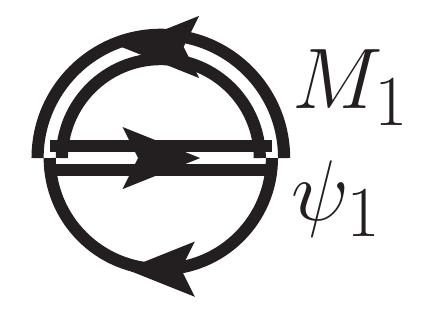}\hspace{-1.5mm}}
  \frac{d_{\psi_2}}{
  \includegraphics[scale=0.3]{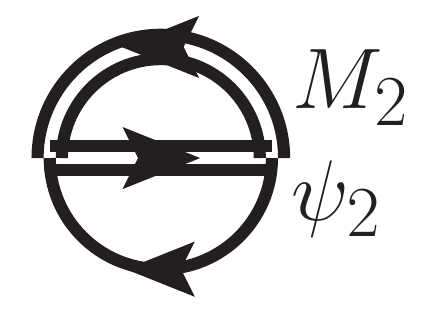}\hspace{-1.5mm}}
  \frac{d_{\psi_3}}{
  \includegraphics[scale=0.3]{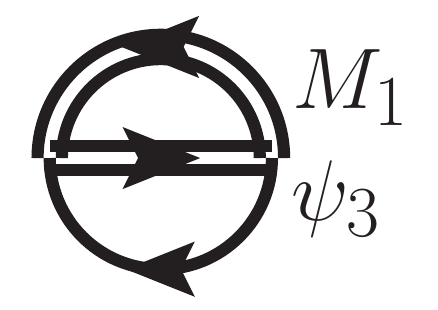}\hspace{-1.5mm}} 
  \raisebox{-0.45\height}{
    \includegraphics[scale=0.45]{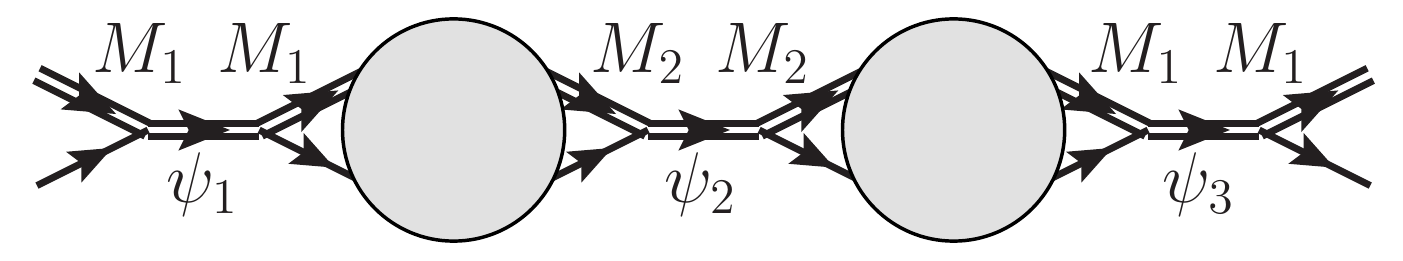}
  }
}\nonumber
\\
& =
  \frac{d_{M'}^2}{
  \bigg(
  \hspace{-1.75mm}
  \raisebox{-0.4\height}{
  	\includegraphics[scale=0.3]{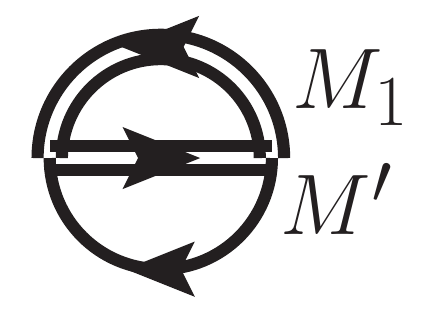}
  }\hspace{-2mm}
  \bigg)^2
  }
  \frac{d_{M'}}{
  \includegraphics[scale=0.3]{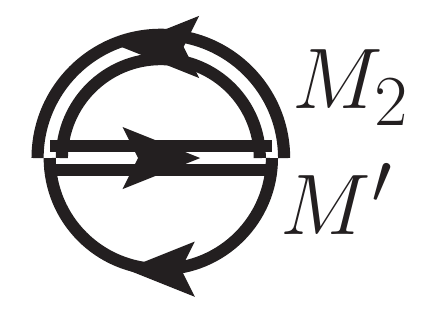}\hspace{-1.5mm}}
  \raisebox{-0.45\height}{
  \includegraphics[scale=0.45]{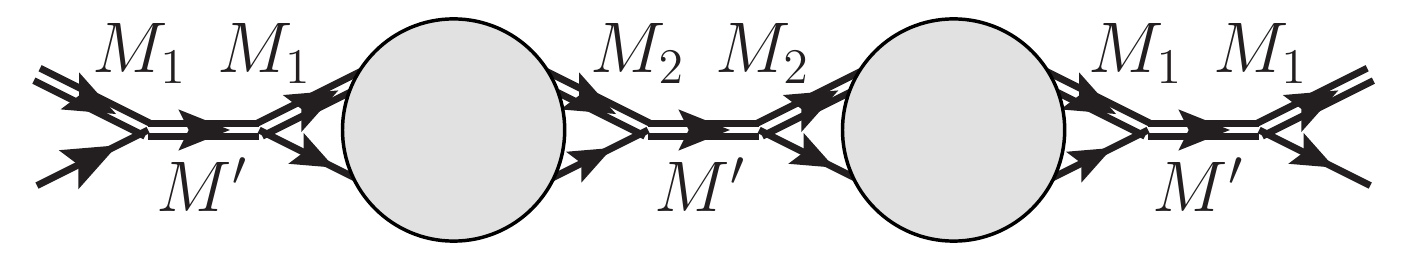}
}\nonumber
\\
& = 
\frac{d_{M'}^3}{
  \bigg(
  \hspace{-1.75mm}
  \raisebox{-0.4\height}{
  	\includegraphics[scale=0.3]{Figures/Two_tensor_products/Wig3jLeftMPrime}
  }\hspace{-2mm}
  \bigg)^2
  \hspace{-1mm}
  \raisebox{-0.4\height}{
  	\includegraphics[scale=0.3]{Figures/Two_tensor_products/Wig3jMiddleMPrime}\hspace{-1.5mm}
  }
}
\frac{
	\raisebox{-0.45\height}{
  		\includegraphics[scale=0.45]{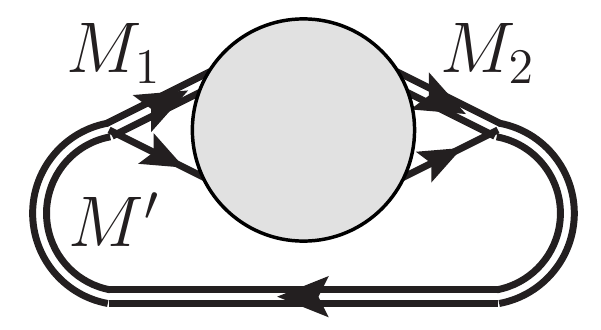}
	}
}{d_{M'}}
\frac{
	\raisebox{-0.45\height}{
  		\includegraphics[scale=0.45]{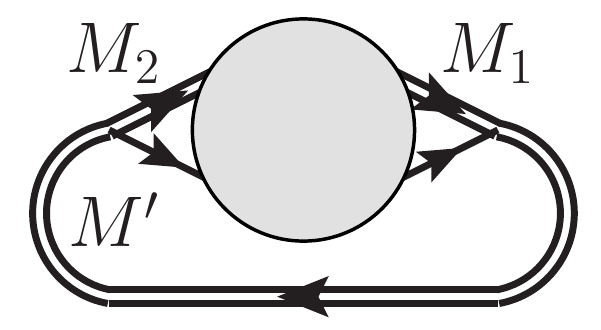}
	}
}{d_{M'}}
\raisebox{-0.45\height}{
  \includegraphics[scale=0.45]{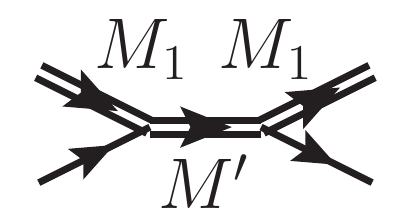}
}\nonumber
\\
&= c P_{M'}.
\end{align}
The gray blobs can again be any connection of lines
between the two color structures
that gives a non-vanishing result. The exact value of the vacuum bubbles containing
the gray blobs does not
matter, as requiring idempotency (or the trace to be equal to the dimension of
the representation) is sufficient to find $c$.
Note that the construction in
\eqref{eq:TwoTensorProductsBirdtrack} gives the projector $P_{M'}$ with the
construction history of $M_1$, whereas the choice of $M_2$ actually is irrelevant
(up to a constant).
Similarly, letting $M_2$ be the outer
representation would give the projector with the construction history of $M_2$.

For finite $\Nc$, it cannot happen that $M'$ exists, but that one choice for
$M_1$ and $M_2$ does not,
since $M_1$ and $M_2$ are given by crossing out one box in $M'$.
Thus we can always construct the needed projectors as indicated in
\eqref{eq:TwoTensorProductsBirdtrack}, as long as the quark diagram in
$M'$ is not rectangular (such that two different representations $M_1$ and $M_2$
cannot be found).
A special case occurs if the antiquark diagram, $\overline{\tilde{Q}}$, correspond to
the singlet representation, and the leftmost column of \eqref{eq:TwoTensorProducts} has
length $\Nc$, since then the first occurrence of the quark diagram is
lowered by $\Nc$ units. For this case we note that the corresponding projector
is still needed, since it corresponds to a valid representation in $M\otimes\V$,
and can still be constructed as for the high $\Nc$ case, using
\eqref{eq:TwoTensorProductsBirdtrack}.

\subsubsection{Construction from one tensor product}
\label{subsec:ProjectorsFromOneTensorProduct}
The remaining representations with higher first occurrence, which cannot be constructed
by the method in \secref{subsec:ProjectorsFromTwoTensorProducts}, all have rectangular
quark diagrams, making it impossible to find two different $M_1$ and $M_2$ as in
\secref{subsec:ProjectorsFromTwoTensorProducts}.
Hence we can consider all representations $M'$ with Young diagram
of form
\begin{equation}
  \label{eq:RectangularRepresentation}
  M' =
  \left(
  \vphantom{\young(\hfil\hfil\hfil\hfil,\hfil\hfil\hfil\hfil,\hfil\hfil\hfil\hfil,\hfil\hfil\hfil)}
  \right.
  \!
  \underbrace{
    \young(\hfil\hfil\hfil\hfil,\hfil\hfil\hfil\hfil,\hfil\hfil\hfil\hfil,\hfil\hfil\hfil\hfil)
  }_{w}
  \scalebox{1.4}{\Bigg\}}h
  ,
  \overline{\tilde{Q}}
  \!
  \left.
  \vphantom{\young(\hfil\hfil\hfil\hfil,\hfil\hfil\hfil\hfil,\hfil\hfil\hfil\hfil,\hfil\hfil\hfil)}
  \right),
\end{equation}
i.e. $h$ boxes high and $w$ boxes wide. For this section there will be three
different cases to deal with, corresponding to different sizes of the quark diagram:
\begin{enumerate}[(i)]
\item The quark diagram is a single box, $Q=\tiny\yng(1)$.
\item The quark diagram consists of two boxes, $Q=\tiny\yng(2)$ or $Q=\tiny\yng(1,1)$.
\item The quark diagram consists of three or more boxes,
  e.g.~$\tiny\yng(4)$, $\tiny\yng(1,1,1)$, $\tiny\yng(2,2)$.
\end{enumerate}

For case (i) and (iii), we consider all representations $M$ with lower first occurrence
than $M'$ that have $M'\in{}M\otimes{}\V$. The only option, in both cases, is the representation
corresponding to the diagram where the bottom right corner of \eqref{eq:RectangularRepresentation}
has been removed (i.e.~the only box in the case of (i)). The tensor product of this representation with $\V$ is
\begin{align}
\label{eq:RectangularYoungTableau}
\left(
\vphantom{\young(\hfil\hfil\hfil\hfil,\hfil\hfil\hfil\hfil,\hfil\hfil\hfil\hfil,\hfil\hfil\hfil)}
\right.
\!
\underbrace{
\young(\hfil\hfil\hfil\hfil,\hfil\hfil\hfil\hfil,\hfil\hfil\hfil\hfil,\hfil\hfil\hfil)
}_{w}
\scalebox{1.4}{\Bigg\}}h,
\overline{\tilde{Q}}
\!
\left.
\vphantom{\young(\hfil\hfil\hfil\hfil,\hfil\hfil\hfil\hfil,\hfil\hfil\hfil\hfil,\hfil\hfil\hfil)}
\right)
\otimes
\yng(1)
&=
\left(
\young(\hfil\hfil\hfil\hfil\hfil,\hfil\hfil\hfil\hfil,\hfil\hfil\hfil\hfil,\hfil\hfil\hfil)
,
\overline{\tilde{Q}}
\right)
\oplus
\underbrace{
\left(
\young(\hfil\hfil\hfil\hfil,\hfil\hfil\hfil\hfil,\hfil\hfil\hfil\hfil,\hfil\hfil\hfil\hfil)
\hspace{0.5mm},
\overline{\tilde{Q}}
\right)
}_{M'}
\oplus
\left(
\young(\hfil\hfil\hfil\hfil,\hfil\hfil\hfil\hfil,\hfil\hfil\hfil\hfil,\hfil\hfil\hfil,\hfil)
\hspace{0.5mm},
\overline{\tilde{Q}}
\right)\nonumber \\ 
&\oplus
\left(
\young(\hfil\hfil\hfil\hfil,\hfil\hfil\hfil\hfil,\hfil\hfil\hfil\hfil,\hfil\hfil\hfil)
\hspace{0.5mm},
\overline{\tilde{Q}}``\otimes{}"\yng(1)
\right).
\end{align}
For case (i), where the quark diagram of $M'$ is $\tiny\yng(1)$,
the quark diagram of $M$ is the singlet, and the first and third terms of
\eqref{eq:RectangularYoungTableau} would not be present.
For case (iii) the projectors corresponding to the first and third terms can be
assumed to have been constructed as in \secref{subsec:ProjectorsFromTwoTensorProducts},
and can thus be projected out. The projectors corresponding to the diagrams
from the last term can also
be projected out, as all contained representations have lower first occurrence and
can be constructed as in \secref{subsec:ProjectorsLowerFirstOccurrence}.
Thus all representations, except the one corresponding to $M'$ can be projected out.
More explicitly, the projector for $M'$ is thus constructed from
\begin{equation}
  \label{eq:OneTensorProductStandardCase}
  P_{M'}
  =
  \raisebox{-0.5\height}{
    \includegraphics[scale=0.45]{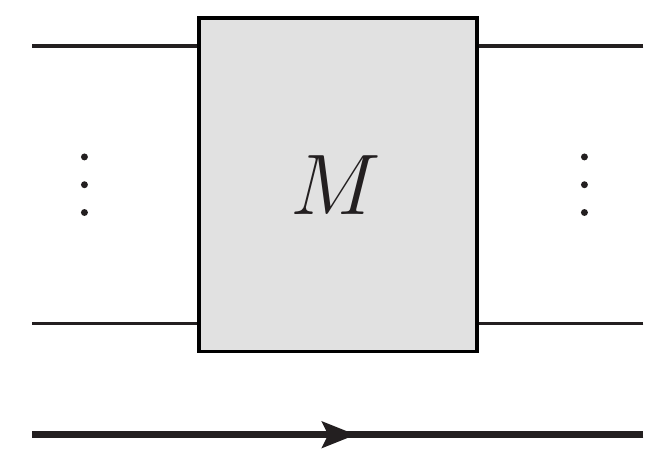}
  }
  -
  \sum_{
    \begin{matrix}
      \scriptstyle
      \psi\in{}M\otimes\,{\tiny \yng(1)}\\
      \scriptstyle
      \psi\neq{}M'
    \end{matrix}
  }{
    \raisebox{-0.5\height}{
      \includegraphics[scale=0.45]{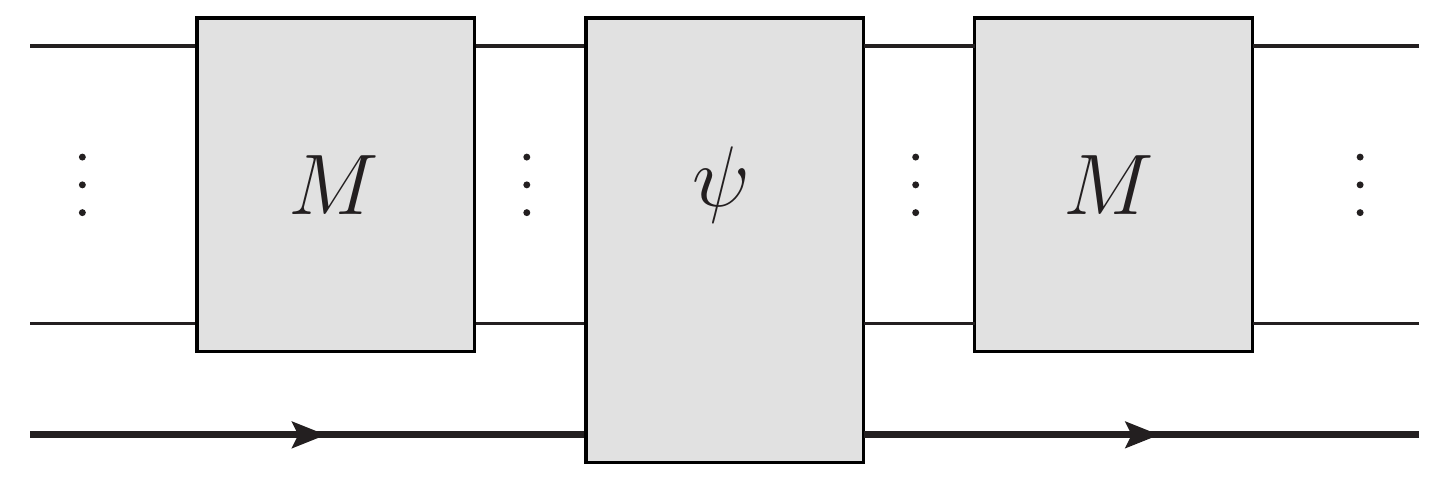}
    }
  }.
\end{equation}

For the final case, (ii), where the quark diagram of $M'$ only has two boxes, the
tensor product corresponding to \eqref{eq:RectangularYoungTableau} is
\begin{equation}
  \label{eq:RectangularYoungTableauSpecialCase}
  \left(
  \yng(1),
  \overline{\tilde{Q}}  
  \right)
  \otimes
  \yng(1)
  =
  \left(
  \yng(2),
  \overline{\tilde{Q}}
  \right)
  \oplus
  \left(
  \yng(1,1)
  \hspace{.5mm},
  \overline{\tilde{Q}}
  \right)
  \oplus
  \left(
  \yng(1)
  \hspace{.5mm},
  \overline{\tilde{Q}}``\otimes{}"\yng(1)
  \right).
\end{equation}
Here we see the issue with this special case, there are two representations
with rectangular quark diagrams, so neither of their projector has,
by assumption, been constructed yet.
As in the previous two cases, the last term corresponds to projectors that can
be constructed by the method in \secref{subsec:ProjectorsLowerFirstOccurrence}.
Thus we construct a tensor
\begin{align}
\label{eq:TSpecialCase}
T
&=
\raisebox{-0.55\height}{
  \includegraphics[scale=0.45]{Figures/SpecialCase/TensorProduct}
}
-
\sum_{
\begin{matrix}
\scriptstyle
M'=({\tiny \yng(1)}\hspace{.3mm},\overline{\tilde{Q}}')\\
\scriptstyle
M'\in{}M\otimes\,{\tiny \yng(1)}
\end{matrix}
}{
\raisebox{-0.55\height}{
  \includegraphics[scale=0.45]{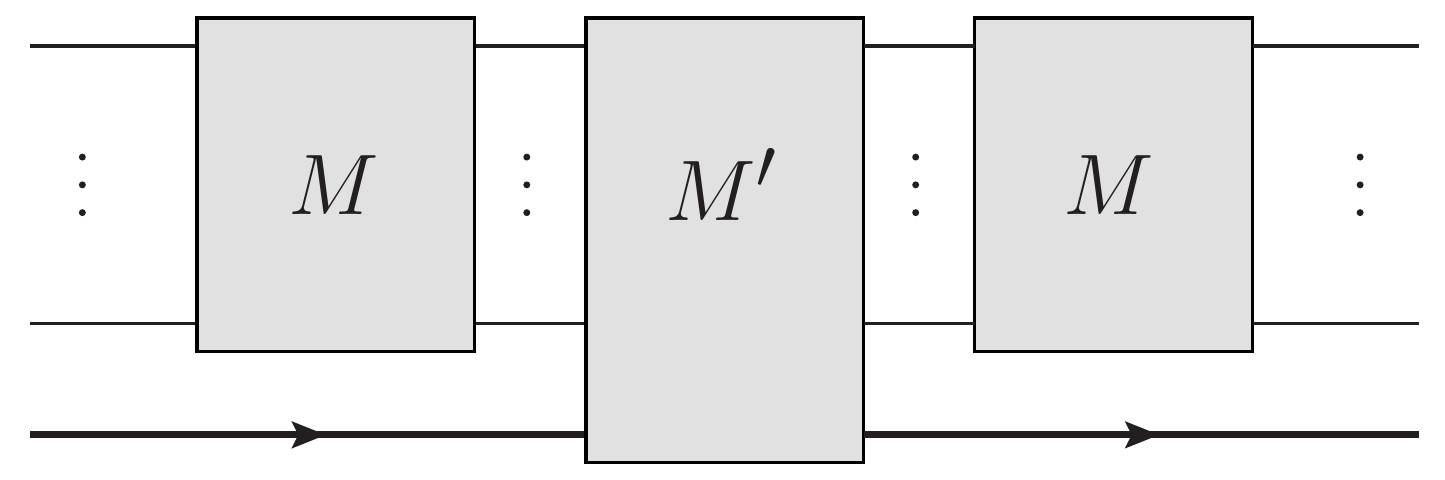}
}
}
\nonumber\\
&=
\raisebox{-0.55\height}{
  \includegraphics[scale=0.45]{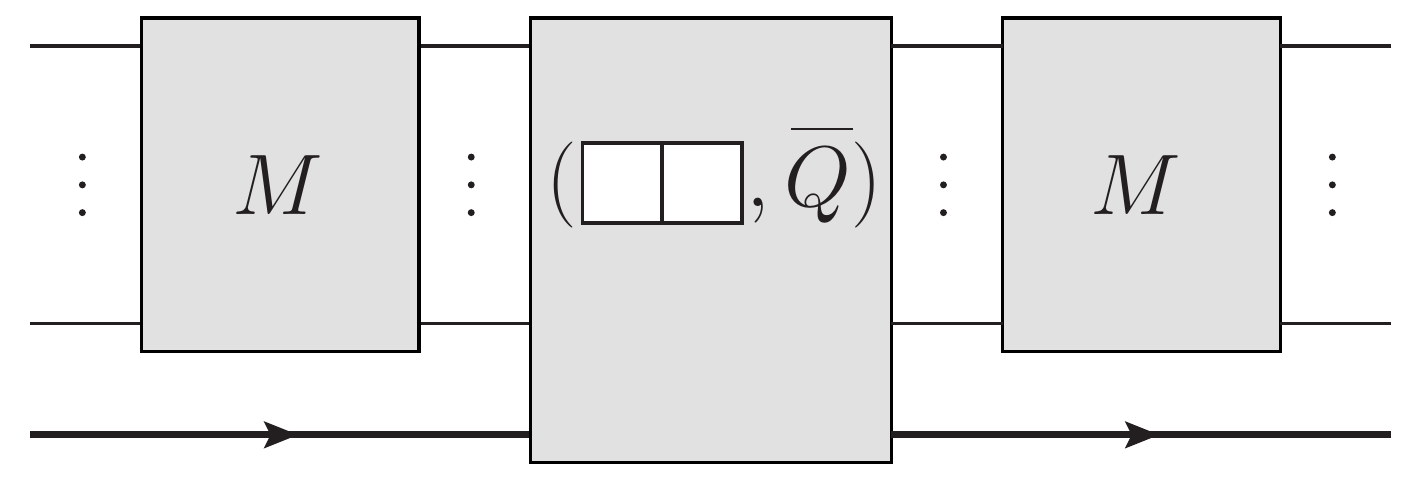}
}
\oplus
\raisebox{-0.55\height}{
  \includegraphics[scale=0.45]{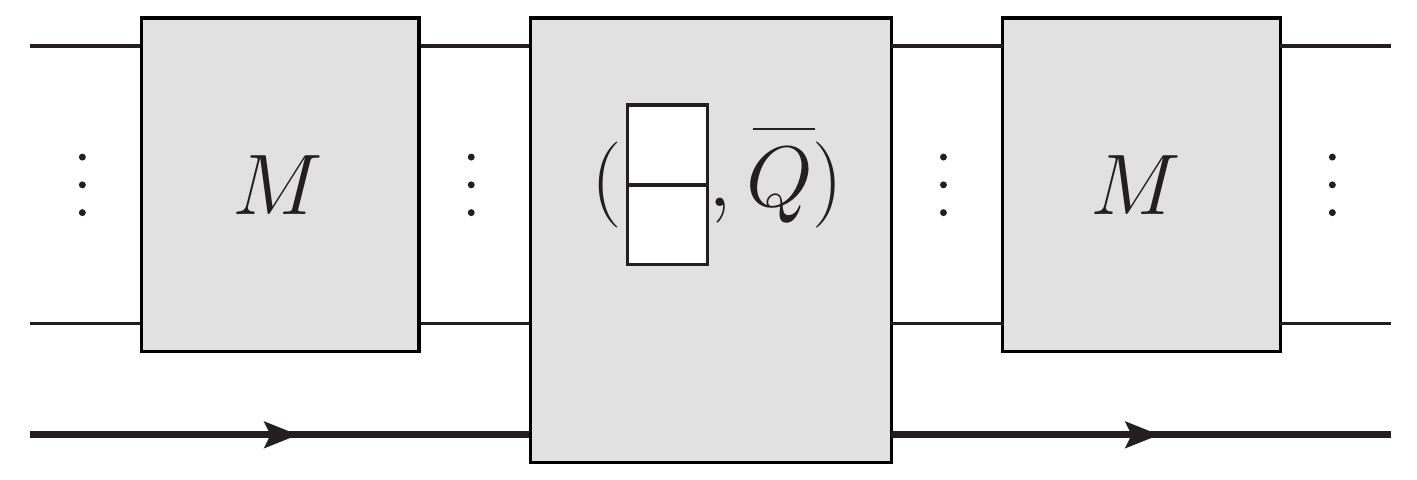}
}
,
\end{align}
where the sum in the first line runs over all the representations coming from the
last term in \eqref{eq:RectangularYoungTableauSpecialCase} for
$M\otimes{}\V$. To pick out one of the remaining two projectors, we can sandwich
a tensor product that does not contain the other projector in between two $T$s.
Such a tensor product is
\begin{equation}
\label{eq:SymmetricTensorProductSpecialCase}
\yng(2)\otimes{}(\cdot,\overline{\tilde{Q}})
=
(\yng(2),\overline{\tilde{Q}})
\oplus
(\yng(1),\overline{\tilde{Q}}')
\oplus\dots\oplus
(\cdot,\overline{\tilde{Q}}'')
\oplus\dots,
\end{equation}
where all the representations, except the first one, have quark diagrams with
one or zero boxes. The overlap of representations in the tensor $T$ and the tensor product
in \eqref{eq:SymmetricTensorProductSpecialCase} is then only
$({\tiny\yng(2)},\overline{\tilde{Q}})$.

We can easily construct the tensor product in
\eqref{eq:SymmetricTensorProductSpecialCase} by using the projector for
$\tiny\yng(2)$, which is simply a symmetrization in the quarks, and the projector
for $(\cdot,\overline{\tilde{Q}})$, which has lower first occurrence, and
hence has been constructed by assumption.
The projector is then proportional to
\begin{equation}
\label{eq:ProjectorSpecialCase}
P_{({\tiny\yng(2)},\overline{Q})} 
\propto
\raisebox{-0.55\height}{
  \includegraphics[scale=0.45]{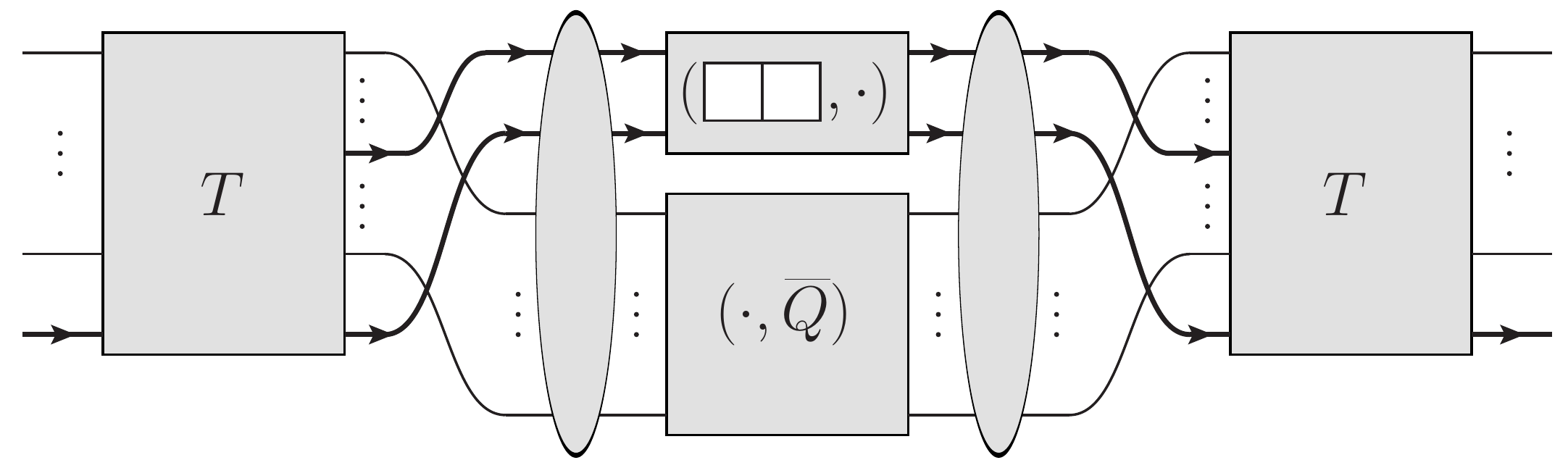}
},
\end{equation}
where again the gray blobs is any connection of lines
that makes the expression non-vanishing.
The projector for the representation
$\left({\tiny\yng(1,1)},\overline{\tilde{Q}}\right)$ can be found in the same manner, with
$\tiny\yng(2)$ changed for $\tiny\yng(1,1)$ in
\eqref{eq:SymmetricTensorProductSpecialCase} and
\eqref{eq:ProjectorSpecialCase}.
It could also be constructed by projecting out $P_{({\tiny\yng(2)},\overline{Q})}$
from $T$.
We thus conclude that we can construct the projectors
also for the special case (ii), where the quark diagram is either
$\tiny\yng(2)$ or $\tiny\yng(1,1)$.
For $\Nc=2$ the quark diagram in the second term in \eqref{eq:RectangularYoungTableauSpecialCase}
is equivalent to the singlet representation, but the corresponding projector can anyway
be constructed as in
\eqref{eq:ProjectorSpecialCase}, using the antisymmetrizer in the middle upper block.

Together with the construction in \secref{subsec:ProjectorsLowerFirstOccurrence},
\eqref{eq:ProjectorLowerFirstOccurrence}, and in \secref{subsec:ProjectorsFromTwoTensorProducts},
\eqref{eq:TwoTensorProductsBirdtrack}, the projectors for any representation $M'$ in the tensor
product of $M\otimes{}\V$ can thus be obtained. The corresponding projectors for $M\otimes{}\Vc$
are constructed analogously.

The projectors constructed by the methods of this section will all
be hermitian (if the previously constructed projectors are hermitian).
In the special case where all lines are incoming quarks (antiquarks) hermitian projectors
can be constructed as in \cite{Keppeler:2013yla,Alcock-Zeilinger:2016sxc}.

\section{Constructing projectors with gluons}
\label{sec:ProjectorsGluon}
The recipe for constructing gluon projectors in \cite{Keppeler:2012ih} is also recursive,
assuming that the projectors with lower first occurrence have already been constructed.
In the tensor product of a representation with $\Adj$, we found in
eqs.~(\ref{eq:GluonCaseSameRep}-\ref{eq:GluonCaseHigherFirstOccurrence}) that the resulting representations
have either lower first occurrence (by one unit of quark first occurrence, and one unit of
antiquark first occurrence), the same first occurrence or higher first occurrence
(one unit of both quark and antiquark first occurrence).
The construction is divided into categories based on how the first occurrence changes.
When the new representation has a lower first occurrence, a construction similar to
\secref{subsec:ProjectorsLowerFirstOccurrence} is employed. For the case of unchanged
first occurrence (\eqref{eq:GluonCaseSameRep} and \eqref{eq:GluonCaseSameFirstOccurrence}),
there is no analogue in this paper, as that cannot occur in the tensor product of a
representation with $\V$. For the case of higher first occurrence, a construction similar
to \eqref{eq:ProjectorSpecialCase} is used, where the quark and the antiquark parts of the
diagram are separated. 

We note that, when constructing the projectors for $M$ in $M\otimes\Adj$,
where there can be more than one occurrence of $M$, the invariance condition
(color conservation)
of color structures can be utilized to find the vertex for one of the
instances of $M$, see section 4.4 in \cite{cvi08}.

\section{Construction of basis vectors from projectors}
\label{sec:basis}
The multiplet basis vectors considered here are of the form
\begin{equation}
  \label{eq:VectorExample}
  \raisebox{-0.45\height}{
    \includegraphics[scale=0.45]{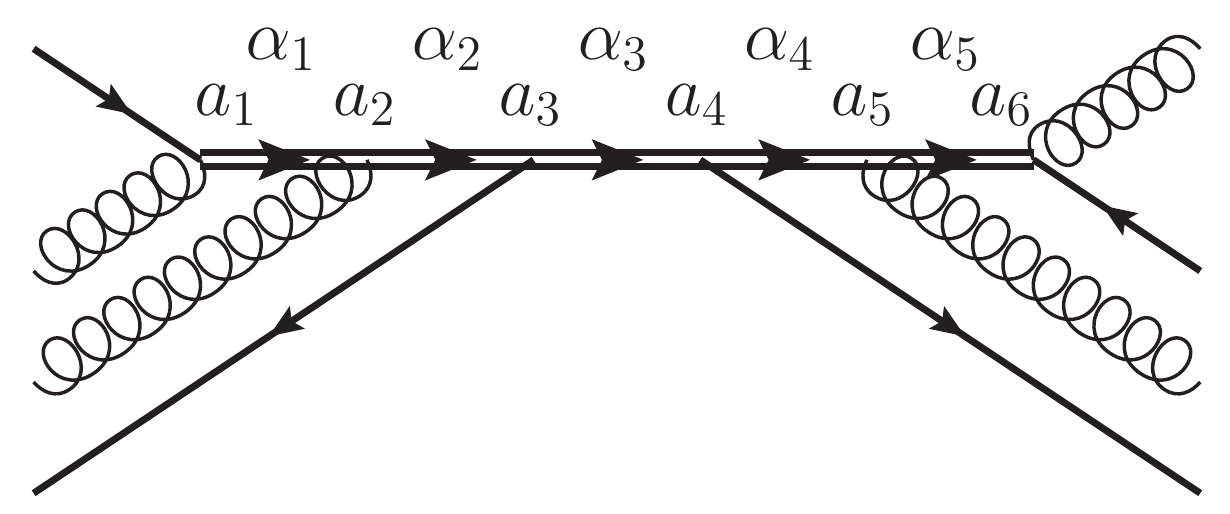}
  },
\end{equation}
where each vector corresponds to a different set of representations $\alpha_i$ and vertices $a_i$.
Recall that the projectors constructed using the method described in
\secref{sec:Projectorsqqbar} and \secref{sec:ProjectorsGluon} can have their partons
in any order. These projectors can then be used to construct basis vectors with the
partons in any order on the left side (e.g.\ in \eqref{eq:VectorExample} the order
is $q$, $g$, $g$, $\overline{q}$) and (in general a different) order on the right
side ($g$, $\overline{q}$, $g$, $q$ in \eqref{eq:VectorExample}). For
perturbative QCD, the vectors we are interested in have the same number of incoming
and outgoing fermion lines. 
The basis vector of \eqref{eq:VectorExample} is proportional to 
\begin{equation}
  \label{eq:VectorExample_Construction}
  \raisebox{-0.45\height}{
    \includegraphics[scale=0.45]{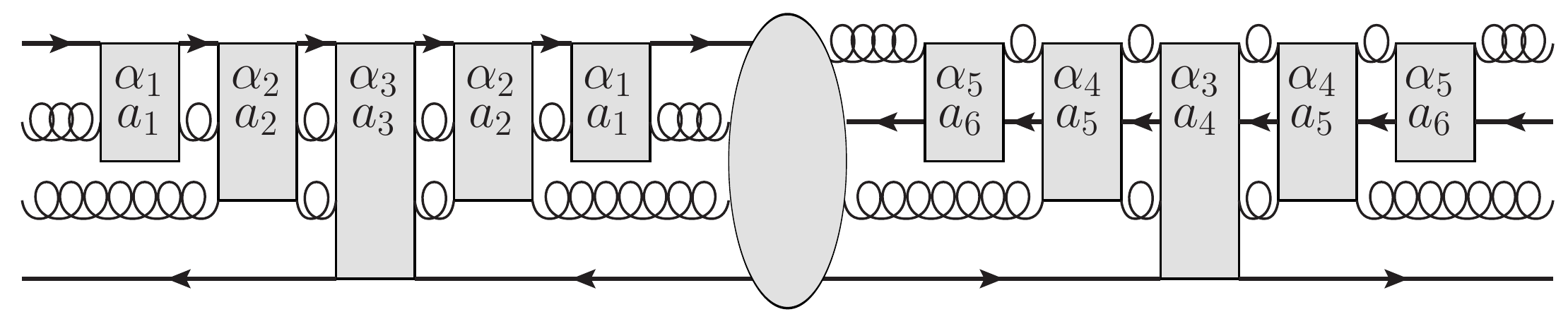}
  },
\end{equation}
where the gray blob in the middle is any non-vanishing connection of lines.
Typically there are many different
possibilities for the gray blob, but they will all give the same color structure up to a
constant, by Schur's lemma, \eqref{eq:SchursLemma}.
The absolute value can be fixed by normalizing the vector,
but the sign of the vector has to be defined carefully (see \secref{sec:signs}).
Contractions of the form in \eqref{eq:VectorExample_Construction} can thus be used to construct
basis vectors for any parton order on the left and right sides. The orthogonality of
the basis vectors follows immediately from the transversality of the projectors (which is why
the basis vectors with the same order of partons on the left and the right sides cannot
have the trivial contraction in the gray blob, unless they are proportional to the
projector).

\subsection{Signs}
\label{sec:signs}
Normally, the sign of a vector in a basis does not matter, since changing the sign of one
vector in a basis still results in a valid basis. For the applications of the bases in this
paper the signs are, however, important.
The basis vectors can be seen as several vertices contracted
with each other, e.g.~\eqref{eq:VectorExample} consists of six vertices,
hence its sign is related to the sign of other vectors
which share some of the vertices.
To consistently use the birdtrack notation, the vectors must also
be related to the mirrored and conjugated vectors,
which can be seen from the completeness relation, \eqref{eq:CRDiagrammatic},
where a vertex always appears with its mirrored and conjugated version.
In the completeness relation, a sign consistency between the two
vertices introduced on the right hand side is required (since every term on the right
hand side is a projector, which has a sign enforced by idempotency). The two
vertices are related by hermitian conjugation, i.e.~conjugating and transposing
one of the vertices gives the other. In birdtracks, vertex conjugation is 
\begin{equation}
  \label{eq:vertex_conjugation}
  \left(\raisebox{-0.4\height}{
    \includegraphics[scale=0.4]{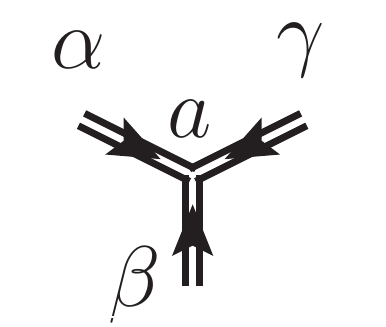}
  }\right)^*
  =\sigma^{\alpha\beta\gamma,a}_c
  \raisebox{-0.4\height}{
    \includegraphics[scale=0.4]{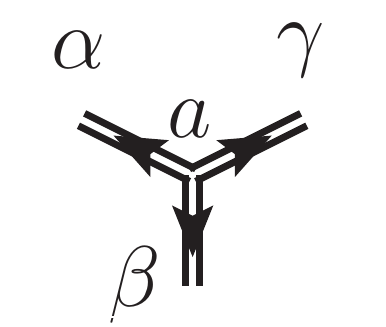}
  },
\end{equation}
and transposition is
\begin{equation}
  \label{eq:vertex_transposition}
  \raisebox{-0.4\height}{
    \includegraphics[scale=0.4]{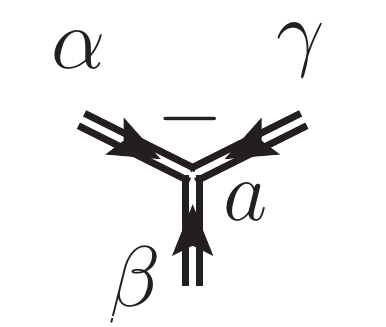}
  }
  \equiv
  \raisebox{-0.4\height}{
    \includegraphics[scale=0.4]{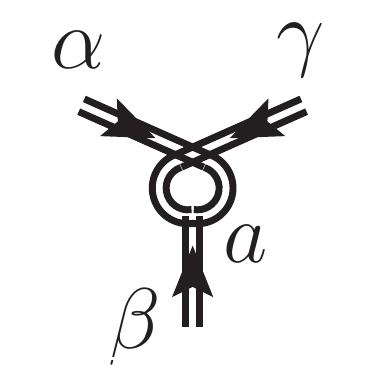}
  }
  =\sigma^{\alpha\beta\gamma,a}_T
  \raisebox{-0.4\height}{
    \includegraphics[scale=0.4]{./Figures/Sign_conventions/Vertex}
  },
\end{equation}
where Yutsis' notation \cite{YutsisNotation} has been used to denote
a vertex in which the representations appear in reversed order.
Hence we require $\sigma^{\alpha\beta\gamma,a}_c\sigma^{\alpha\beta\gamma,a}_T=1$ for all
vertices such that the completeness relation is fulfilled. This sign convention
is also the reason that the
scalar product of two color structures can be depicted by contracting
one of the color
structures with the mirror image of the other color structure with its arrows
reversed, as in \figref{fig:ScalarProduct}.
The signs under conjugation $\sigma^{\alpha\beta\gamma,a}_c$ and
transposition $\sigma^{\alpha\beta\gamma,a}_T$ can be chosen freely if neither of
\eqref{eq:vertex_conjugation} and \eqref{eq:vertex_transposition}
has the same vertex on the left hand and right hand sides,
since we can always redefine one
of the vertices to be minus itself (the vertex related by hermitian conjugation
would then also have to change sign)\footnote{
  This would have the effect of changing the sign of some of the basis vectors
  and Wigner $6j$ coefficients. The scalar product between a basis vector
  and a color structure would also only change by at most a sign, if the
  basis vector changed sign. As the decomposition in \secref{sec:using_wigner6j}
  only introduces new vertices in the completeness relation \eqref{eq:CRDiagrammatic}
  and in \eqref{eq:VertexCorrection}, and they always come with a vertex and its
  hermitian conjugate, the decomposition will remain unaffected.
}.
In \eqref{eq:vertex_conjugation} the two vertices are the same if
all three representations are real, in which case the sign is easily determined
by conjugating the vertex. For \eqref{eq:vertex_transposition} the two vertices are the same
if, at least, two of the representations are identical.

In the electronic appendix we have used the freedom of choice to set
$\sigma^{\alpha\beta\gamma,a}_{c/T}=1$, when possible.

\section{Calculation of $6j$ coefficients}
\label{sec:6j}
From the basis vectors in \secref{sec:basis} we can evaluate scalar products
as described in \cite{Sjodahl:2015qoa}, to obtain all the Wigner $6j$ coefficients
in \eqref{eq:Wigner6j3Representations} and \eqref{eq:Wigner6j4Representations}.
The coefficients in \eqref{eq:Wigner6j4Representations} can be found by evaluating
\begin{equation}\label{eq:Wigner6j4repCalculation}
\Tr{
\left[
\raisebox{-0.4\height}{
\includegraphics[scale=0.4]{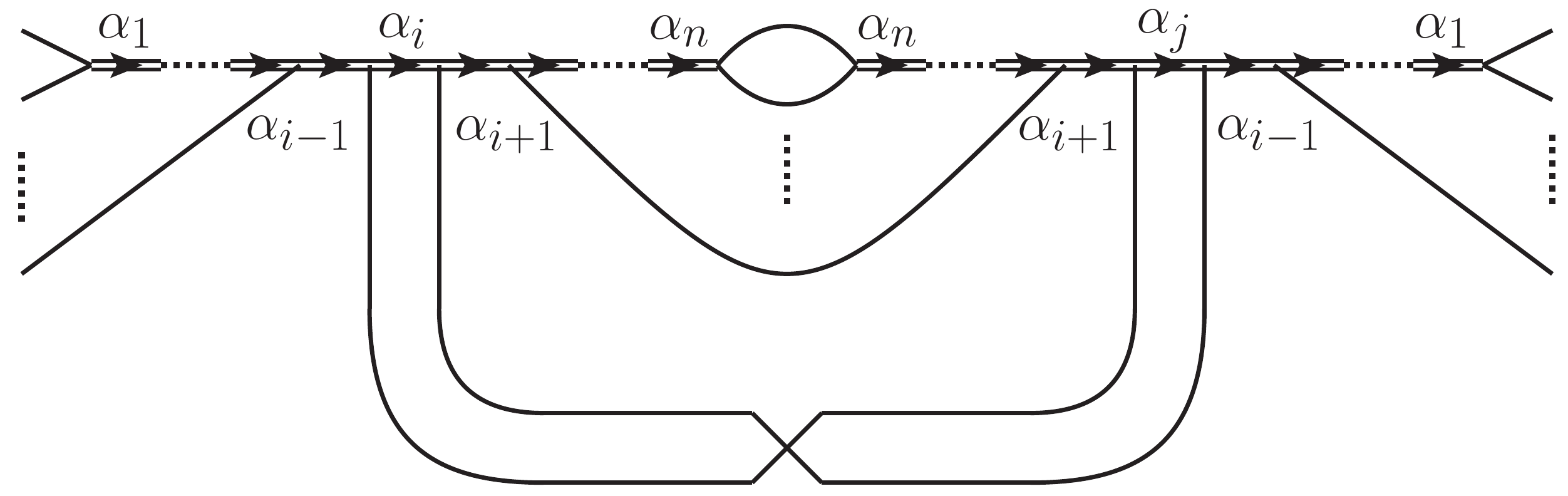}
}
\right],
}
\end{equation}
where, in the basis vectors, all representations except the representation at
position $i$ are identical.
This can be simplified, using Schur's lemma, \eqref{eq:SchursLemma}, to remove
the two-vertex loops, into a contraction of the form
\begin{equation}
  \label{eq:Wigner6j4repCalculationStep1}
=
\prod_{k=1}^{i-1}{
	\frac{
		\raisebox{-0.45\height}{
		\includegraphics[scale=0.4]{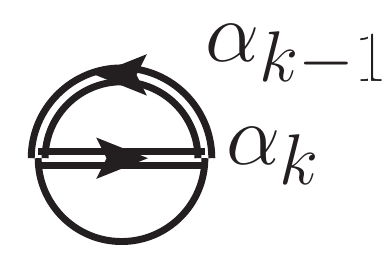}
		}
		\hspace{-2.5mm}
	}{
	d_{\alpha_k}
	}
}
\prod_{l=i+1}^{n}{
  \frac{
    \raisebox{-0.45\height}{
      \includegraphics[scale=0.4]{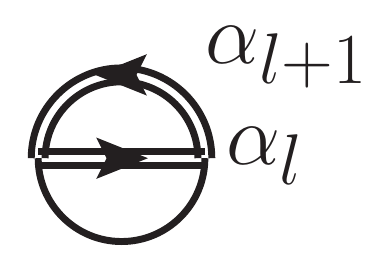}
    }
    \hspace{-2.5mm}
  }{
    d_{\alpha_l}
  }
}
\raisebox{-0.4\height}{
  \includegraphics[scale=0.4]{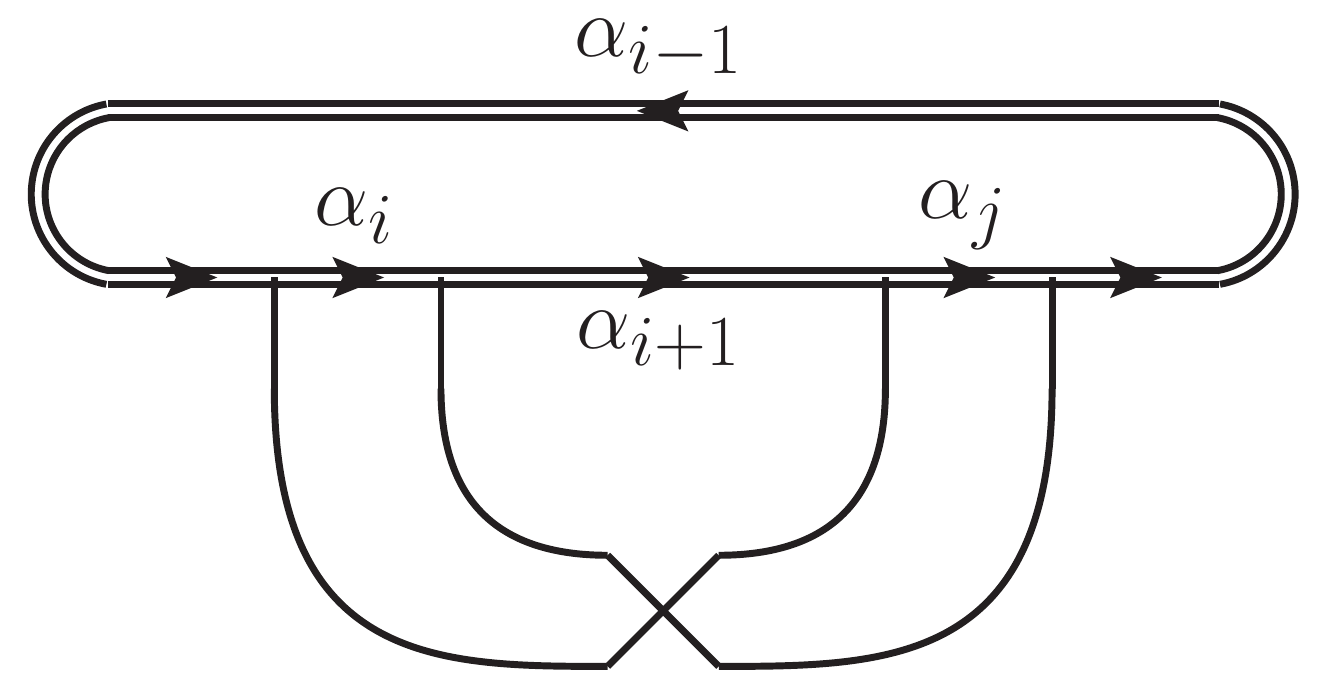}
},
\end{equation}
where $n$ is the number of representations in the basis vectors and $\alpha_0$
(which occurs in the first product) is the representation of the first leg (to the top
left in \eqref{eq:Wigner6j4repCalculation}) and $\alpha_{n+1}$ is the representation of the
leg in the top middle of \eqref{eq:Wigner6j4repCalculation}.
We can construct basis
vectors such that in the contraction corresponding to \eqref{eq:Wigner6j4repCalculation}
the crossed lines can be in any of the representations $\V$, $\Vc$ or $\Adj$.
The Wigner $3j$ coefficients are
just vertex normalization factors, and can be set to, for example, unity.
In this paper all Wigner $3j$ coefficients have been kept, hence all equations are valid
for any choice of vertex normalization. In the electronic appendix however,
a specific normalization has been chosen, where all vertices are normalized
(including the $ggg$ and $q\overline{q}g$ vertices) such that all non-vanishing
$3j$ coefficients have value one. 

As the prefactors in \eqref{eq:Wigner6j4repCalculationStep1} are trivial to
calculate, we will drop them now, and consider what happens if we replace the crossed
lines in \eqref{eq:Wigner6j4repCalculation} with $\V$, $\Vc$ and $\Adj$. If both of the lines
are gluons we have
\begin{equation}
  \label{eq:Wig6j4rep_gg}
  \raisebox{-0.4\height}{
    \includegraphics[scale=0.4]{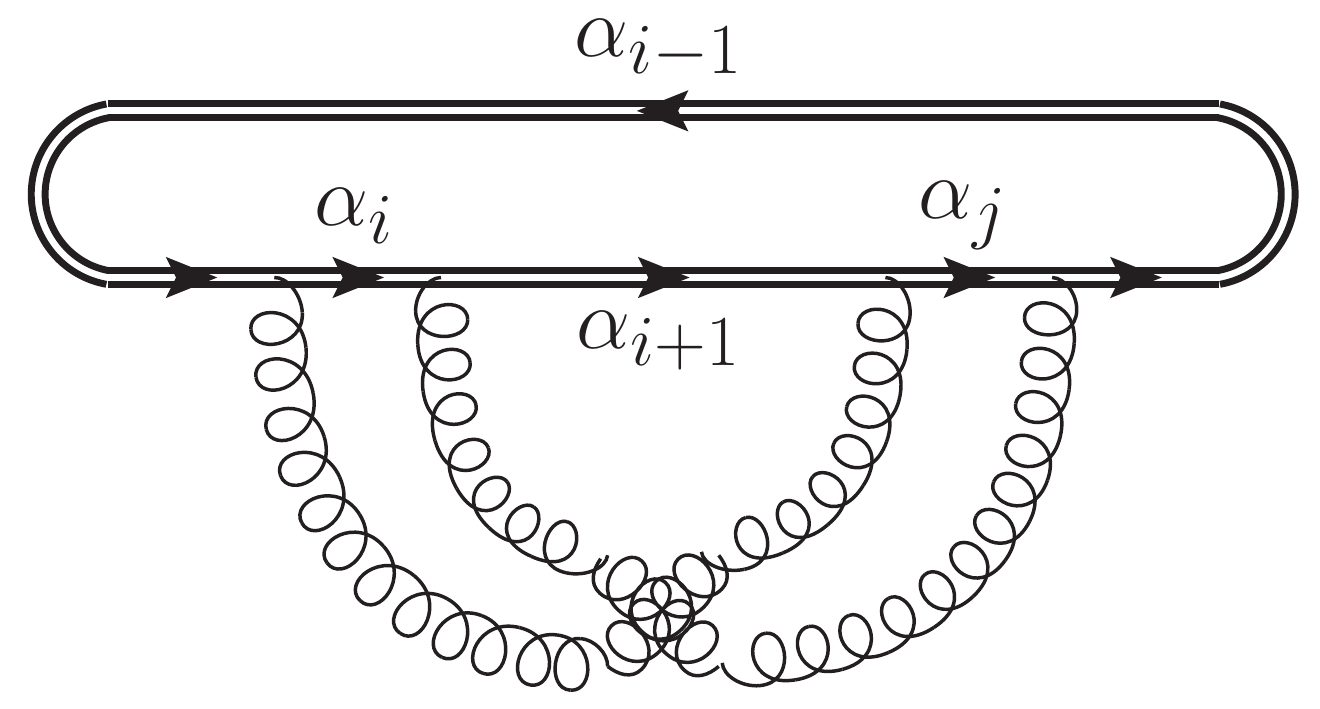}
  }
  =
  \raisebox{-0.4\height}{
    \includegraphics[scale=0.4]{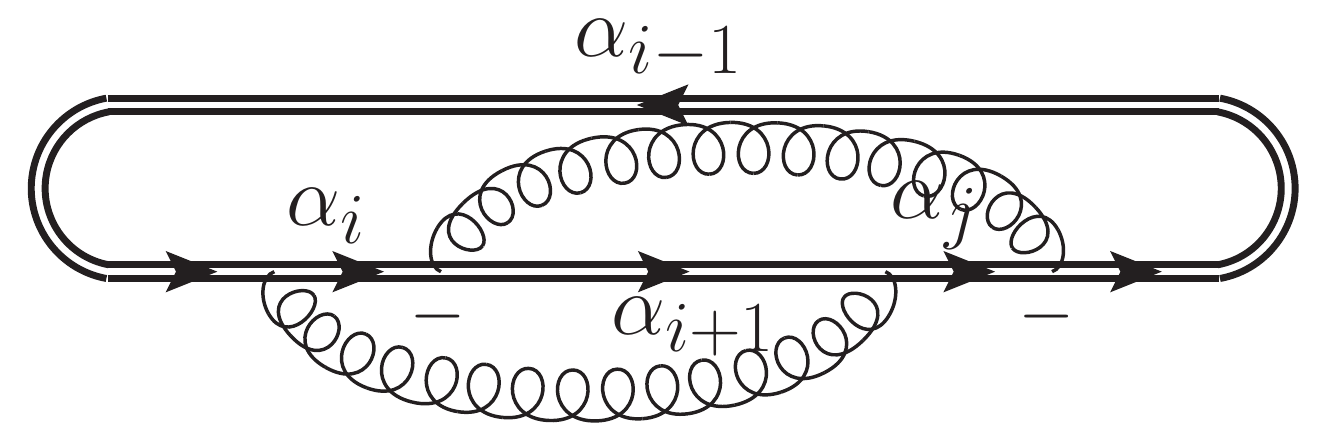}
  }
  =
  \raisebox{-0.4\height}{
    \includegraphics[scale=0.4]{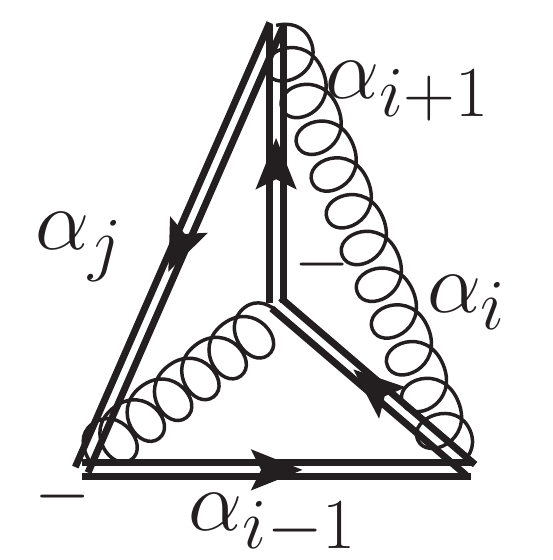}
  },
\end{equation}
i.e.\ the first $6j$ coefficient in \eqref{eq:Wigner6j4Representations} (up to two vertex
orderings).
If one of the lines is a quark, we get
\begin{equation}
  \label{eq:Wig6j4rep_qg}
  \raisebox{-0.4\height}{
    \includegraphics[scale=0.4]{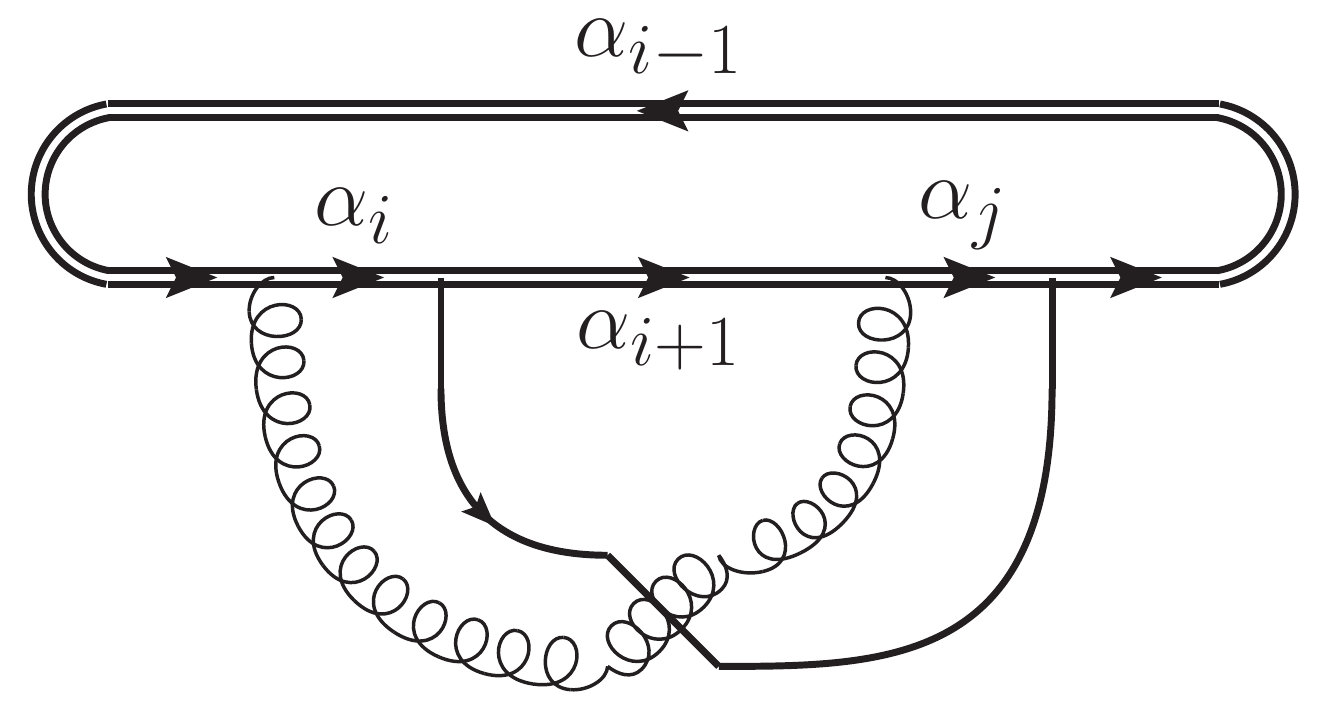}
  }
  =
  \raisebox{-0.4\height}{
    \includegraphics[scale=0.4]{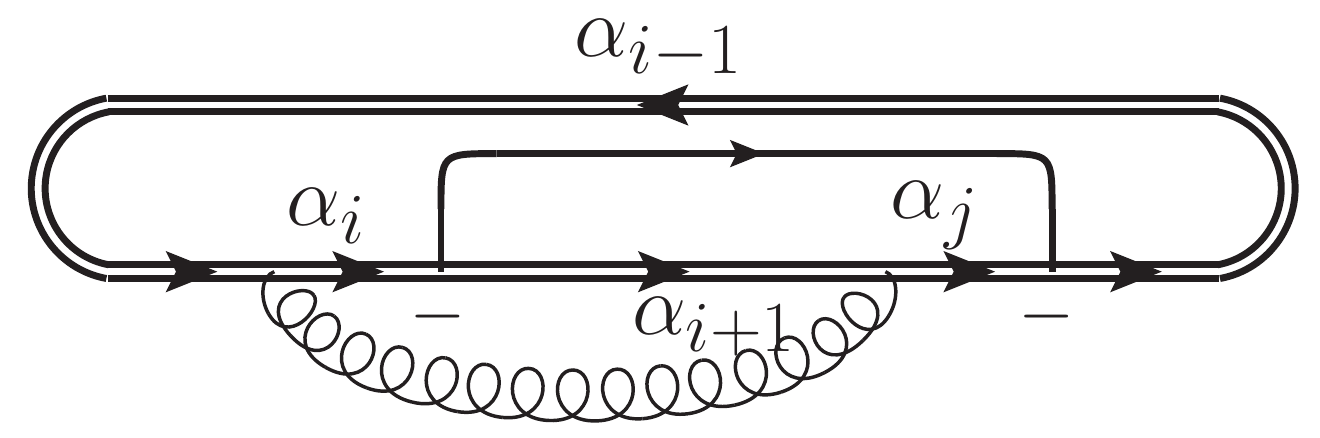}
  }
  =
  \raisebox{-0.4\height}{
    \includegraphics[scale=0.4]{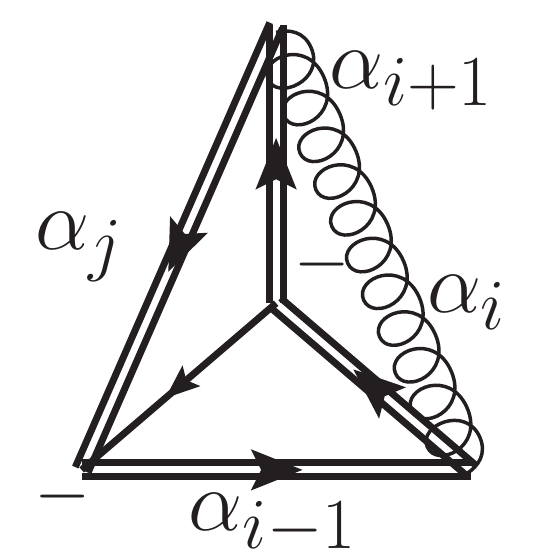}
  },
\end{equation}
corresponding to the middle coefficient in \eqref{eq:Wigner6j4Representations}.
Finally, the last coefficient in \eqref{eq:Wigner6j4Representations} is obtained
if both of the lines are quarks,
\begin{equation}
  \label{eq:Wig6j4rep_qq}
  \raisebox{-0.4\height}{
    \includegraphics[scale=0.4]{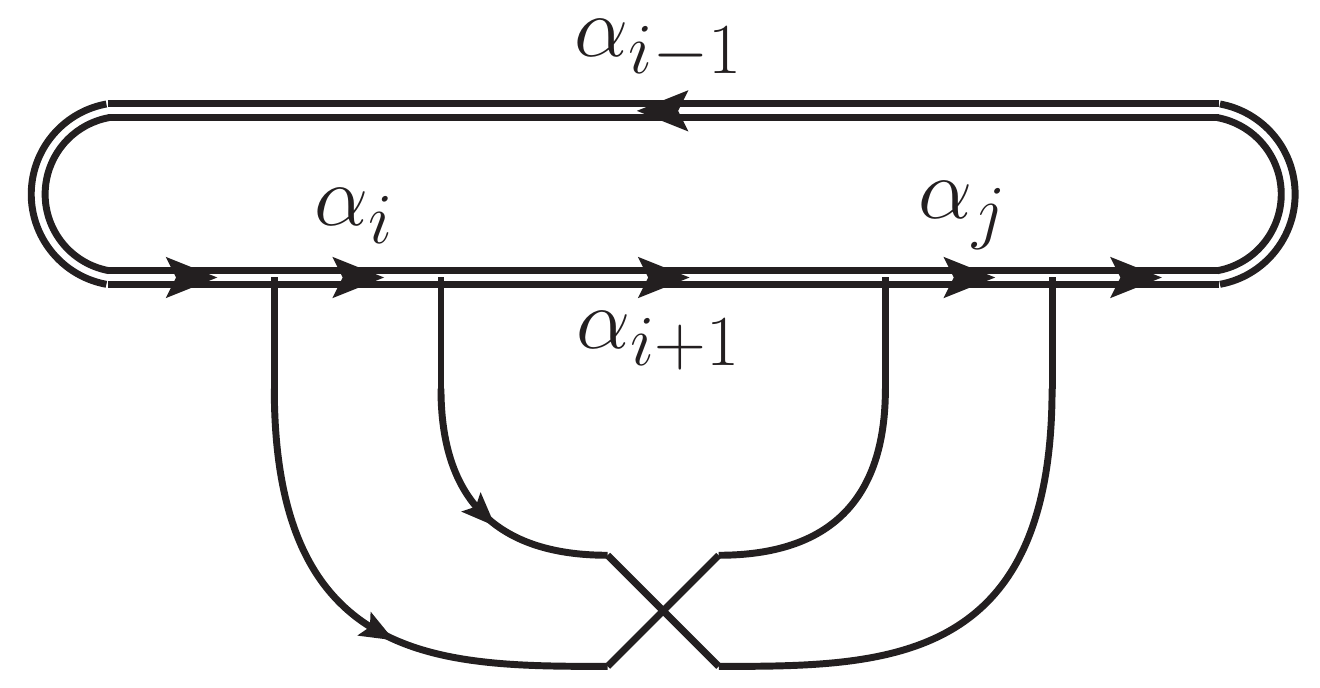}
  }
  =
  \raisebox{-0.4\height}{
    \includegraphics[scale=0.4]{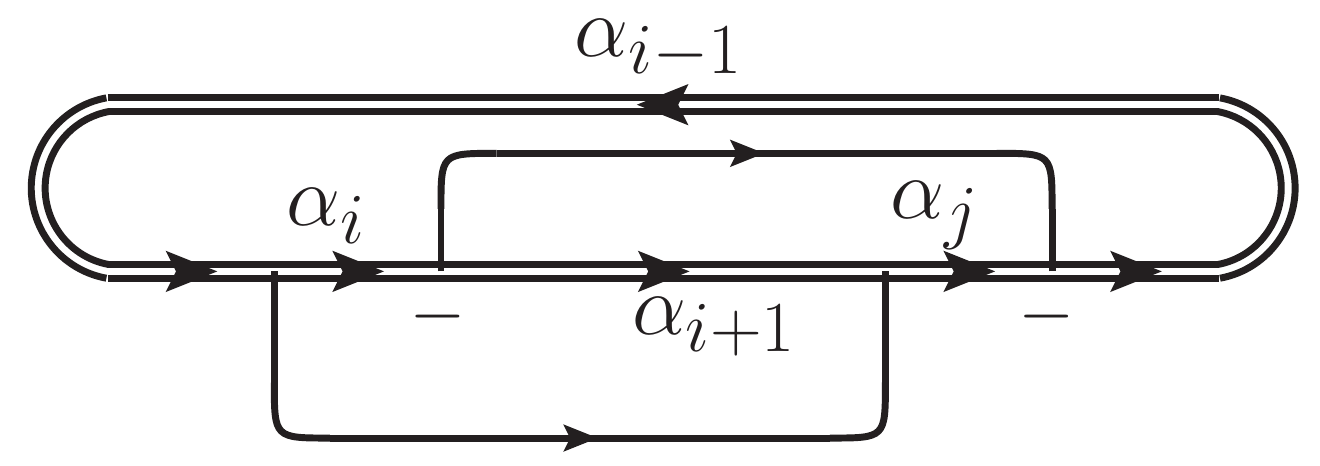}
  }
  =
  \raisebox{-0.4\height}{
    \includegraphics[scale=0.4]{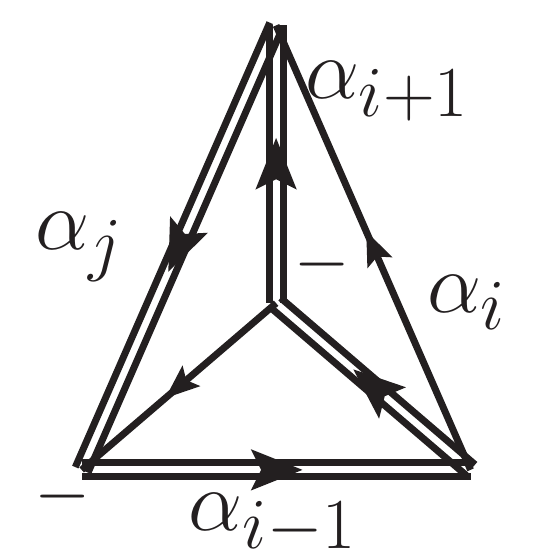}
  }.
\end{equation}
Due to symmetries of the coefficients, the cases of antiquark
lines instead of quark lines can always be related to the quark cases. Thus all Wigner
$6j$ coefficients in \eqref{eq:Wigner6j4Representations} can be calculated.

The coefficients in \eqref{eq:Wigner6j3Representations} are calculated similarly,
but using one basis vector for $N_p$ partons and one for $N_p+1$ partons,
\bea\label{eq:Wigner6j3repCalculation}
& &
\Tr{
\left[
\raisebox{-0.4\height}{
\includegraphics[scale=0.4]{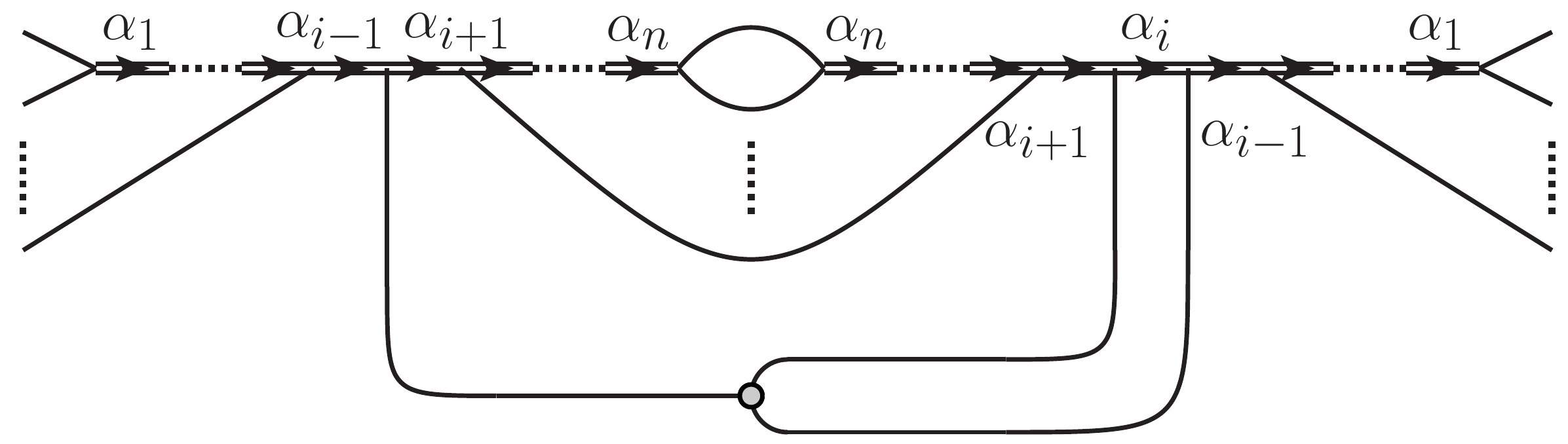}
}
\right]
}
\nn
&&=
\prod_{k=1}^{i-1}{
	\frac{
		\raisebox{-0.45\height}{
		\includegraphics[scale=0.4]{Figures/WignerEvaluation/WigProducts/Wig3j_LeftRight}
		}
		\hspace{-2.5mm}
	}{
	d_{\alpha_k}
	}
}
\prod_{l=i+1}^{n}{
	\frac{
		\raisebox{-0.45\height}{
		\includegraphics[scale=0.4]{Figures/WignerEvaluation/WigProducts/Wig3j_Middle}
		}
		\hspace{-2.5mm}
	}{
	d_{\alpha_l}
	}
}
\raisebox{-0.4\height}{
\includegraphics[scale=0.4]{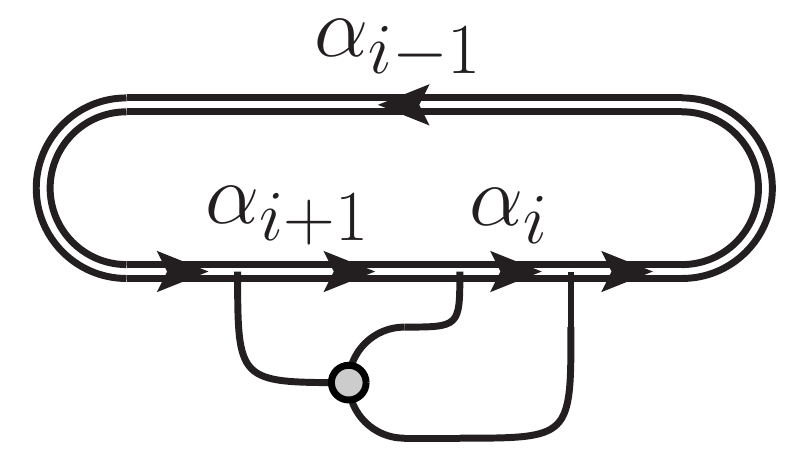}
},
\eea
where the gray blob can be the antisymmetric triple gluon vertex or the
fermion-gluon vertex. Again we drop the trivial prefactor and only consider the rightmost
vacuum bubble of \eqref{eq:Wigner6j3repCalculation}. If the gray blob is the antisymmetric
triple gluon vertex we get
\begin{equation}
  \label{eq:Wig6j3Rep_f}
  \raisebox{-0.4\height}{
    \includegraphics[scale=0.4]{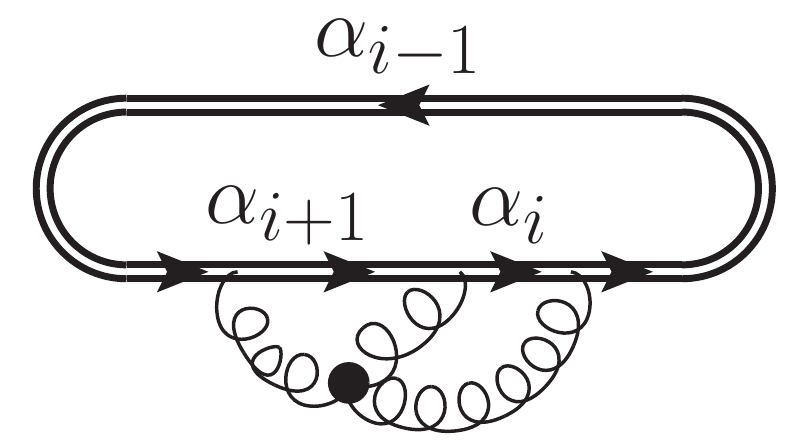}
  }
  =
  \raisebox{-0.4\height}{
    \includegraphics[scale=0.4]{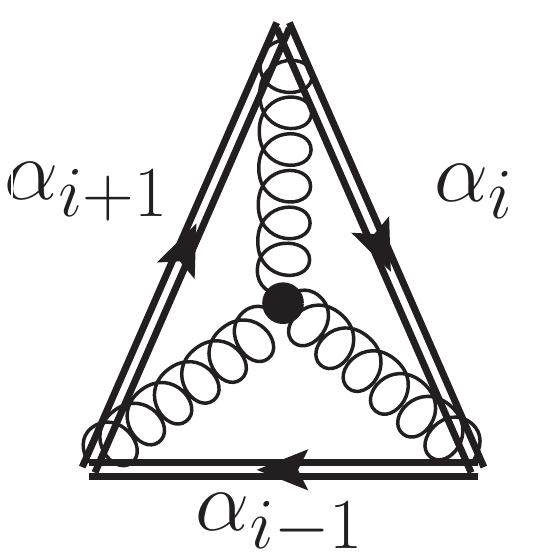}
  },
\end{equation}
corresponding to the first coefficient in \eqref{eq:Wigner6j3Representations}.
The second coefficient of \eqref{eq:Wigner6j3Representations}, then corresponds to the
case where two of the representations entering the gray blob are in the quark and
antiquark representations,
\begin{equation}
  \label{eq:Wig6j3Rep_t}
  \raisebox{-0.4\height}{
    \includegraphics[scale=0.4]{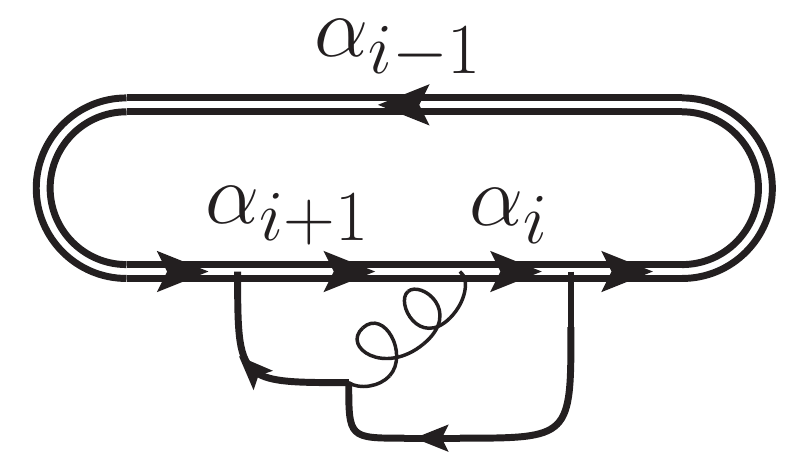}
  }
  =
  \raisebox{-0.4\height}{
    \includegraphics[scale=0.4]{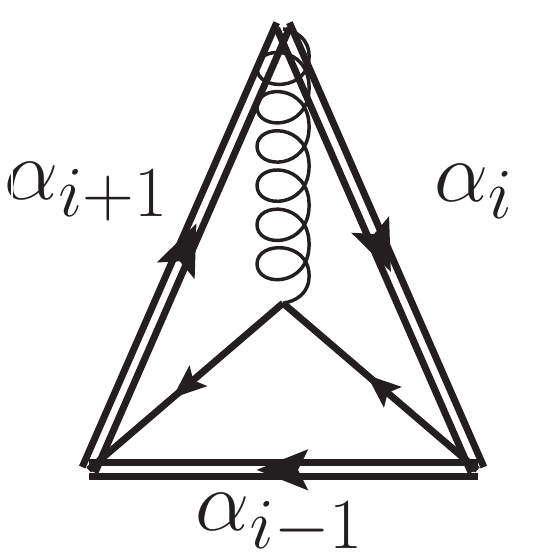}
  }.
\end{equation}
Hence, by calculating contractions of the form \eqref{eq:Wigner6j4repCalculation}
and \eqref{eq:Wigner6j3repCalculation}, all of the coefficients in \eqref{eq:Wigner6j4Representations} and \eqref{eq:Wigner6j3Representations} can be calculated.

Since the Wigner $6j$ coefficients have many symmetries, we only need to calculate a few of them.
This has been done in the electronic appendix for up to four external quark-antiquark pairs or
gluons, meaning that projectors and vectors with up to four incoming and outgoing quarks
or antiquarks have been constructed along with the $6j$ coefficients in
\eqref{eq:Wigner6j4Representations} and \eqref{eq:Wigner6j3Representations}.
The calculations are performed using ColorMath \cite{Sjodahl:2012nk}, and the resulting
Wigner 6j coefficients are attached (in a human readable format) in Wigner6jCoefficientsWithQuarks.m.

\section{Conclusions}
\label{sec:conclusions}
This paper introduces an algorithm for constructing $SU(\Nc)$ projectors onto representations
corresponding to quarks, antiquarks and gluons, grouped in any order
onto a chain of backbone representations (see \figref{fig:basis_comparison}). These projectors
can then be used to construct multiplet bases, and from the bases, Wigner $6j$ coefficients can be
calculated. The Wigner coefficients allow for an efficient decomposition of color structures
into multiplet bases.

As the more general projectors and basis vectors introduced in this paper remove
constraints on groupings of partons for decompositions involving quarks,
they are better suited for any application where
the grouping matters, compared to the basis vectors in \cite{Keppeler:2012ih}.
One such case is amplitude decomposition into multiplet bases
for amplitudes with quarks.
The method for using Wigner $6j$ coefficients, summarized in \secref{sec:using_wigner6j},
is more efficient if shorter loops can be used
and for color structures with quarks,
this is enabled with the Wigner $6j$ coefficients
constructed here, as compared to the method presented in \cite{Sjodahl:2015qoa}.

The grouping of partons also matter for amplitude recursion relations with quarks,
since the efficiency of the recursion depends on how similar the $N_p$ parton multiplet
basis is to the $N_p+1$ parton
basis \cite{Du:2015apa}. Loosening the requirement from \cite{Keppeler:2012ih}
of grouping the quarks and antiquarks into singlets or octets,
bases giving fewer terms in the recursion step can be chosen.
As parton showers have similar color structures to recursion relations,
due to the fact that one parton at the time is added,
the bases in this paper could also be used for $\Nc=3$ parton showers. 
As for the recursion relations in \cite{Du:2015apa}, this should give a
very significant speed up compared to using trace or color flow
bases for full color treatment.

\acknowledgments
We thank Stefan Keppeler for useful comments on the manuscript.
This work was supported
by the Swedish Research Council (contract numbers 2012-02744 
and 2016-05996), 
and in
part by the MCnetITN3 H2020 Marie Curie Initial Training Network,
contract number 722104, as well as the European Union's Horizon 2020
research and innovation programme (grant agreement No 668679).

\bibliographystyle{JHEP}  
\bibliography{Refs} 

\end{document}